**A hybrid predictive and prescriptive modelling framework for long-term mental healthcare workforce planning**

Harsha Halgamuwe Hewage[a,1], Bahman Rostami-Tabar[a,2]

[a]Data Lab for Social Good Research Group, Cardiff Business School, Cardiff University, Aberconway Building, Colum Drive, Cardiff CF10 3EU, United Kingdom

[1]halagmuwehewagehr@cardiff.ac.uk, [2]rostami-tabarb@cardiff.ac.uk

**Abstract**

Over the past decade, severe staffing shortages in mental healthcare have worsened due to rising demand, further exacerbated by COVID-19. This demand is expected to grow over the next decade, necessitating proactive workforce planning to ensure sustainable service delivery. Despite its critical importance, the literature lacks a comprehensive model to address long-term workforce needs in mental healthcare. Additionally, our discussions with UK NHS mental health practitioners highlight the practical need for such a model. To bridge this gap, we propose a hybrid predictive-prescriptive modelling framework that integrates long-term probabilistic forecasting with an analytical stock-flow model for mental health workforce planning. Given the pivotal role of nurses, who comprise one-third of the mental health workforce, we focus on forecasting nursing headcount while ensuring the model's adaptability to broader healthcare workforce planning. Using statistical and machine learning methods with real-world NHS data, we first identify key factors influencing workforce variations, develop a long-term forecasting model, and integrate it into an analytical stock-flow framework for policy analysis. Our findings reveal the unsustainable trajectory of current staffing plans and highlight the ineffectiveness of blanket policies, emphasizing the need for region-specific workforce strategies.

## 1 Introduction

Over the last decade, the growth of the mental health workforce has been limited. For instance, many essential staff groups have either remained at the same level or even declined since 2009 in the UK (HEE, 2022). Among these segments, mental healthcare nurses constitute a vital component, making up a third of the entire mental healthcare workforce (Palmer et al., 2023). This underscores the crucial role that nurses play in the mental healthcare system. Nearly one-third of all nursing vacancies within NHS England pertain to mental healthcare nurses, even though they comprise only about 13% of the total nursing workforce (British Medical Association, 2019), revealing a pressing need for a comprehensive and sustainable strategy to address the workforce shortage.

To ensure a skilled, suitable, and enduring nurse workforce for the mental healthcare sector in the future, a long-term workforce strategy is essential (HEE, 2022). Such a strategy should be developed using workforce planning models that consider essential factors to adapt to the evolving mental healthcare needs of the population. Consequently, the primary goal of workforce planning should be to strike the right balance between the supply and demand of the mental healthcare nursing workforce in both the short and long terms (Ono et al., 2013). Neglecting proper planning poses the risk of shortages or surpluses, which can have a negative impact on patient care quality and resource efficiency (Hovenga & Lowe, 2020).

Our investigation holds significance both theoretically and practically, as decision-makers have access to an ever-expanding pool of healthcare data and a growing need to understand how data can enhance workforce planning. Moreover, there is limited evidence on the most appropriate driving factors, including both supply and demand, in the literature for nurse workforce planning (Weller-Newton et al., 2021). Therefore, our study addresses a clear practical need for mental healthcare workforce planning, which was highlighted during our discussions with key mental health practitioners working for the NHS in the UK. Furthermore, Willis et al. (2018) highlight that it is crucial to capture the existing complexities in workforce planning and incorporate them into the design of OR models for workforce projection methods. However, Willis et al. (2018) explain that we cannot accurately forecast or measure the accuracy of long-term forecasts in healthcare workforce planning due to many exogenous factors that we as modellers are unable to control, such as the healthcare needs of people, technological developments, or unforeseen events like a pandemic. Thus, healthcare workforce planning must consider a multi-method approach that can incorporate dynamic complexities into the model (Tako & Kotiadis, 2015). This requires the development of integrated predictive and prescriptive modelling approaches, which have so far received relatively limited attention in the literature related to healthcare workforce planning. While some studies have separately employed these approaches in the context of healthcare, no prior research has focused on the development of a model to inform long-term workforce needs to ensure effective mental healthcare planning by combining forecasting and simulation, despite the topic's significance (Willis et al., 2018).

While the integration of forecasting and simulation has been previously explored in healthcare workforce planning, our study makes a novel contribution by applying this hybrid predictive–prescriptive framework to the under-researched domain of mental health nursing. Specifically, we combine a long-term probabilistic forecasting model with an analytical stock-flow simulation, using a uniquely comprehensive dataset drawn from multiple national sources. This enables us to uncover key demand and supply-side drivers and evaluate policy interventions at a regional level, offering a robust and context-specific decision-support tool for mental health workforce planning.

Hence, our research offers three distinct contributions, namely:

1. We propose a hybrid predictive and prescriptive modelling framework by combining probabilistic forecasts with a simulation model using the analytical stock-flow method. This allows for the determination of shortfalls and surpluses in the workforce under different demand and supply scenarios and offers a comprehensive policy decision analysis.
2. We identify driving factors influencing long-term staffing needs and develop long-term probabilistic forecasting models to forecast future headcounts for a horizon of six years.
3. We propose an empirically grounded approach to health workforce planning by combining an uncommon number of data sources, including NHS Workforce Statistics, the NHS Mental Health Dashboard, OECD health data, Mental Health Services Monthly Statistics, NHS Sickness Absence Rates, NHS Vacancy Statistics, NMC Registration Data Reports, and Population and Migration Statistics. This thorough collection makes a substantial contribution to the development of models for future mental health workforce planning.

This research complies with the principles of reproducibility. We provide the R and Python codes for forecasting models, as well as the stock-flow method developed in Anylogic. While the main focus of our research is on the nursing mental health workforce, it is essential to emphasise the adaptability of the suggested framework, as it may be used in many other healthcare services, given the nature of the problem and the availability of data.

The structure of our paper is as follows: Section 2 discusses the relevant literature and establishes the theoretical foundation for the research problem. The methodology is presented in Section 3. Section 4 comprises a comprehensive analysis of the empirical study's findings. Finally, Section 5 focuses on discussion and conclusion.

## 2 Literature Review

Operational research (OR) methods have been extensively applied to improve the decision-making process in healthcare settings. A review by Long & Meadows (2018) found that existing OR literature on mental healthcare services mainly focuses on healthcare system operations, healthcare system design and planning, treatment evaluation, and medical decision-making. Moreover, Long & Meadows (2018) and Noorain et al. (2019) state that Markov modelling, system dynamics (SD), agent-based modelling (ABM), and discrete event simulation (DES) are the widely applied OR methods in mental healthcare services. However, Noorain et al. (2023) highlights that workforce planning in mental healthcare has received little attention in the literature. Unlike workforce planning methods in other healthcare services, which consider factors related to service delivery like skill mix, work patterns, and healthcare service utilisation and account for uncertainty, such developments are still lacking in mental healthcare (Noorain et al., 2023).

However, the sustainability of a healthcare system relies on the dependability of its workforce. Therefore, workforce planning centres on ensuring an appropriate number of staff, equipped with the necessary skills, are situated in the correct locations at the right times to provide necessary services to those seeking healthcare (Sutton et al., 2023). In simpler terms, workforce planning is the technical process of forecasting care demand and the essential staff to deliver that care. The primary purpose of healthcare workforce planning is to strike a balance between the supply and demand of the healthcare workforce, both in the short and long terms (Ono et al., 2013). This is significant because a shortage of healthcare staff in any service can lead to poor services, while a surplus can result in the wastage of resources (Fattahi et al., 2023).

Long-term workforce planning mechanisms enable decision-makers to gain insight into future workforce requirements and implement strategies to achieve improved healthcare goals and objectives (Benahmed et al., 2023). These goals and objectives are often interconnected with key factors such as economic policies, healthcare education, rules and guidelines for healthcare workforce development, and meeting population health and social care needs (Carter & Busby, 2023). For example, effective workforce planning can help decision-makers make policy-level decisions regarding nursing education enrolments based on available budget constraints (Ono et al., 2013). Conversely, poor workforce planning can have significant negative impacts. It puts people's lives at risk, may increase avoidable medical errors, elevate morbidity rates, and waste taxpayer money due to misallocations that result in economic inefficiencies (Asamani et al., 2018).

However, healthcare workforce planning can be challenging for three key reasons. Firstly, there can be systematic delays from both the supply and demand sides. For instance, it takes 3 to 4 years to train a nurse, but during this period, changes in required skills might occur due to technological innovations. On the demand side, changes in population behaviour, such as smoking, can affect healthcare needs (Willis et al., 2018). Secondly, the combinatorial complexity arises from the availability of numerous healthcare services, leading to various healthcare paths. This results in different service delivery configurations that require different workforce skill mixes (Weller-Newton et al., 2021). Thirdly, dynamic complexities emerge due to different workforce behaviours. For instance, foreign-trained workers, attrition rates, or internal transfers add complexity to workforce planning (Buerhaus et al., 2015). Therefore, it's crucial to consider all these aspects when developing an effective healthcare workforce planning approach.

## 2.1 Workforce planning in healthcare

The planning of the healthcare workforce involves both qualitative and quantitative methods, with qualitative data often being utilized to model contextual factors such as expert opinions or policy decisions (Sutton et al., 2023). All these methods can be categorized into three approaches: 1) supply-based, 2) demand-based, and 3) need-based (Willis et al., 2018).

Supply-based models often focus on balancing the healthcare workforce inflow and outflow (Sutton et al., 2023). Thus, they concentrate on determining the number of healthcare workers at a given time based on factors such as joiners, graduates, training, retirement, and transfers (Safarishahrbijari, 2018). They provide an answer to the question, "How many healthcare workers will there be?" On the other hand, the demand-based approach centres on the current and future demand for healthcare services to identify the required number of healthcare workers (Willis et al., 2018). This approach addresses the query, "How many healthcare workers will we need?" Additionally, the need-based approach strives to provide the target level of service based on population health needs (Safarishahrbijari, 2018). It seeks to answer, "How many healthcare workers do we need to provide a certain level of service?"

### 2.2 Qualitative and Quantitative Methods

Qualitative methods are commonly used in supply-based and need-based approaches, while quantitative methods dominate demand-based approaches (Willis et al., 2018). Qualitative methods often rely on expert judgment in workforce planning, using techniques like surveys or the Delphi method (Safarishahrbijari, 2018). These models are particularly important when no empirical data is available (Hovenga & Lowe, 2020). Nevertheless, there are criticisms of qualitative methods, including their reliance on experts, which can introduce bias into the planning process. Consequently, workforce projections may lack accuracy, and the process may prove time consuming (Ono et al., 2013).

Given the need to consider both supply and demand factors alongside population needs, quantitative based models have emerged as a viable alternative to qualitative methods. These models centre around OR methods like time series analysis, regression, simulation, and optimization (Asamani et al., 2018).

Time series models are widely used in healthcare workforce planning applications (Sutton et al., 2023). Exponential smoothing, autoregressive integrated moving average (ARIMA), Markov modelling, and vector error correction models are popular implementations in time series models (Safarishahrbijari, 2018). These methods rely on historical data and provide projections based on trends and seasonal features (Malgieri et al., 2015). Although these models consider patterns of past data, their main disadvantage is assuming those patterns will continue in the future (Ono et al., 2013). Also, they are unable to incorporate exogenous features such as attrition rates, wage increments, or special periods like a pandemic (Safarishahrbijari, 2018). However, these approaches provide a simple and efficient way to make workforce predictions despite these limitations.

On the other hand, regression-based models overcome the limitations of time series models by incorporating exogenous variables into their predictions. Consequently, they can identify causal relationships between the target variable and the predictors (McLean & McIntosh, 2021). However, these methods can be considered sophisticated when numerous predictor variables are involved, and they necessitate expert guidance for model creation and maintenance (Kolassa et al., 2023). Another constraint is the requirement for both historical data and future values of the predictors (Willis et al., 2018).

Simulation is a useful method for assessing future healthcare workforce scenarios based on the potential impacts of various changes in demand and supply factors (Sutton et al., 2023). Thus, simulation models are commonly used in healthcare workforce modelling. These models can be categorized into 1) discrete event simulation, 2) agent-based modelling, 3) system dynamics (SD), and 4) hybrid modelling, a combination of these models (Preiss et al., 2022). Optimization methods, on the other hand, have also been used in healthcare workforce planning. The common methods include linear programming, mixed integer linear programming, and dynamic programming (Howick & Ackermann, 2011). These approaches, however, are often complex and necessitate expert guidance for successful implementation.

Each of these models offers distinct dimensions to workforce planning. For example, literature shows that in recent years, modelers have recognized the need to apply time series or a regression model to take into account stochastic features while using simulation models like SD with expert knowledge to model the dynamic nature of the workforce (Willis et al., 2018). Thus, it gives a holistic view of the future workforce to decision makers.

## 2.3 Related works

Buerhaus et al. (2015) proposed a workforce projection model for registered nurses in the United States utilizing data from the U.S. Census Bureau. They employed a straightforward statistical model to estimate age and cohort effects for each U.S. region. These cohort effects, along with population projections, were then used for making workforce projections. Nevertheless, this model failed to incorporate potential changes in healthcare policies or other factors that might impact the nursing workforce. Additionally, they assumed that future cohorts entering the workforce would follow the same patterns observed in recent cohorts, which might not be a valid assumption. Conversely, a similar study conducted in Belgium utilized the Delphi method in combination with a simple statistical approach (Kroezen et al., 2018). This study found that incorporating the Delphi method improved the forecasting process by quantifying the uncertainty of input variables. In the work by Marco et al. (2021), a mathematical approach was utilized to estimate the baseline workforce supply of physicians, nurse practitioners, and physician assistants. Growth rates were then calculated for making projections. Scenario analysis was also employed to assess potential outcomes based on assumptions about workforce supply and demand trends. Furthermore, Bienemy (2015) developed an Excel-based mathematical model to project nursing figures regionally. This model integrated both supply and demand side projections, including sensitivity analysis based on feature variables.

Moreover, Benahmed et al. (2023) employed both quantile regression for demand-side projection and stock and flow analysis for supply-side projection in their workforce projection model for obstetrics and gynaecology. They also conducted scenario analysis based on training programs. In the works of Crettenden et al. (2014), they used a dynamic stock and flow model to estimate future healthcare workforce supply in Australia. They considered workforce-related data such as new joiners, migrations, and net exits, as well as demand-related data such as healthcare utilization rates. They conducted different scenario analyses focusing particularly on nurses, involving productivity improvements and retention rate changes. Gupta and Basu (2007) similarly employed a stock and flow model for nurse supply projection in Canada, integrating it with a utilization-based approach for demand-side factors. Also, Murphy et al. (2012) introduced a need-based approach using a stock and flow model. Their supply side for nurses was developed through the stock and flow approach. For determining present and future needs, factors like population-based considerations, healthcare utilization levels, productivity, and service level targets were employed.

Santric-Milicevic et al. (2013) adopted time series methods including ARIMA and transfer function models to forecast physician and nurse supply in Serbia. The model considered variables such as population size, GDP value, inpatient care discharges, outpatient care visits, students enrolled in medical studies, and graduated physicians. Additionally, McRae (2021) utilized ARIMA, exponential smoothing, regression, and neural networks to predict regional demand for hospital services. Moreover, Asamani et al. (2018) employed the Markov process to estimate future demand for healthcare services. This projection formed the basis for estimating healthcare workforce requirements using service standards.

Li et al. (2007) described a Multiple Objective Linear Programming (MOLP) model for estimating the demand for nurses. The primary focus of this model is to determine the requirement for hiring, training, and promoting staff in the upcoming years. On the other hand, Lavieri & Puterman (2009) utilized linear programming (LP) to develop a hierarchical workforce planning model for nurses in British Columbia, Canada. They performed what-if analysis to create scenarios and identify the optimal recruitment plan. Additionally, Schell et al. (2016) employed an LP model to estimate the need for neonatal nurse workforce, along with a system dynamics model to identify the optimum policy suggestions. Moreover, Hu et al. (2016) proposed an infinite linear programming model to identify the optimal need for nurse recruitment in British Columbia. They conducted scenario analysis and natural look-ahead policy analysis to address the uncertainty of the future requirements.

However, reviews by Ono et al. (2013) and Safarishahrbijari (2018) found that many healthcare workforce planning models often focus on either estimating future supply or the demand for workforce, but not both. Table 1 also shows that, from the reviewed literature, nurse workforce planning methods often focus on only one aspect. A smaller number of approaches concentrate on the need-based approach. While several approaches rely on simple equation-based mathematical models, attempts have been made to integrate different methods to create effective workforce planning models. In contrast, our proposed method incorporates both supply and demand by integrating a machine learning-based probabilistic forecasting model with an analytical stock-flow model. Thus, our model has the potential to account for uncertainty in both the demand and supply sides.

Table 1: A comparison of relevant previous studies with our proposed model.

| Source | Approach[1] | Coverage | Supply factors | Demand factors | Other factors | Projection period | Models | Uncertainty methods[2] |
|---|---|---|---|---|---|---|---|---|
| Ono et al. (2013)[3] | Supply based and demand based simulation approach[3] | Physicians, nurses and midwives | Number of graduates headcount, FTE, exit rates, age distribution | Population, utilisation rate | None | 2010-2025 | Stock-flow model | Scenario analysis |
| Ono et al. (2013) | Supply and demand based simulation approach[4] | Physicians and nurses | Number of graduates, foreign staff inflow headcount, FTE, retirement | Population, utilisation rate | None | 2000-2025 | Stock-flow model | None |
| Ono et al. (2013) | Need based simulation approach[5] | Registered nurses | Education, immigration, retirement, death, out-migration, number of license holders', participation rates, activity rates, level of service | Population, utilisation rate | None | 2007-2022 | Stock-flow model | Scenario analysis |
| Ono et al. (2013) | Supply based simulation approach[6] | Nurses | Training capacity, training quota, nurse graduates' headcounts, exit rates, transfers | Population | None | 2006-2030 | Micro-simulation model | Scenario analysis |
| Ono et al. (2013) | Supply based simulation approach[7] | Nurses | Nursing graduate's headcount, exit rate, career changes | Population, utilisation rate | None | 2005-2025 | Stock-flow model | Scenario analysis |
| Buerhaus et al. (2015) | Supply based statistical approach | Nurses | Age distribution, employment status, working hours | Population | Geographical area | 2010-2030 | Simple statistical model | None |
| Marco et al. (2021) | Supply and demand based forecasting model | Physicians, nurses, physician assistants | Headcount, activity rate, age distribution, training | Population, utilisation rate | None | 2024-2030 | Mathematical model | Sensitivity analysis |
| Bienemy (2015) | Supply and demand based forecasting model | Nurses | Headcount, age distribution, work patterns | Population, age, distribution, utilisation rate, nurse to patient ratio, degree of urbanization, healthcare service -demand | Geographical area | - | Excel-based mathematical model | Sensitivity analysis, feedback from experts |
| Santric-Milicevic et al. (2013) | Supply based forecasting approach | Physician and nurses | Student enrolments, number of graduates, headcount, productivity rate | Population, inpatient care discharges, outpatient care visits | GDP | 2009-2015 | ARIMA and TF models | Sensitivity analysis |
| Li et al. (2007) | Demand based optimisation approach. | Nurses | Job type and category, service hours, overtimes, salary level | Scheduled patient visits, unscheduled patient visits | Costs associate with admin tasks | - | MOLP | - |
| Lavieri & Puterman (2009) | Supply based optimisation approach | Nurses | Number of students, number of nurses and managers recruits, number of nurses promoted, number of managers promoted, age, number of nurses, managers and senior managers | Population, utilisation rate | Costs associate with admin tasks | 2007 - 2027 | LP | What-if analysis |
| Schell et al. (2016) | Need based optimisation and simulation approach | Neonatal nurses | Sizes of nursing programs, nursing workforce size, proportion of student categories, exam passing rate, attrition rate | Annual demand growth for nurses | - | 2014 - 2034 | LP and SD | Scenario analysis |
| Hu et al. (2016) | Supply and demand based optimisation approach | Nurses | Number of nurses, workforce levels, retention rate, attrition rate, number of students, promotions | Demand for nurses | Costs associate with admin tasks | - | Infinite linear programming | Sensitivity analysis, scenario analysis |
| **Proposed Model** | Supply based simulation and demand based forecasting approach | Mental healthcare nurses | Leaver headcounts, joiners' headcounts, absence rate, nurse graduates, foreign trained nurses, vacancy rates, UCAS acceptance size, UCAS acceptance rate | Number of referrals, number of people in contact with mental health services, number of people with an open hospital spell, number of contacts Population | Expenditure on mental healthcare | 2023-2028 | sNAIVE, ARIMA, ETS, linear regression, gradient boosting models, stock-flow model | Probabilistic forecast, what-if analysis, feedback from experts |

[1] We categorized each paper based on the main objective of the model.
[2] Considering the variety of terms in the literature that describe how a model's output reacts to changes in its inputs, we interchangeably use sensitivity analysis, scenario analysis, and what-if analysis in this paper.
[3] This review study, conducted by the OECD in 18 countries, examined 26 models. Among these, we only included methodologies related to nurse workforce projections; [3]Australia: Health Workforce 2025 plan, [4]Canada: Health Human Resource Model, [5]Need-Based Simulation Model for Registered Nurses in Canada, [6]Demographic Projection of Nurses 2030, [7]Forecasting Nursing Professions

## 3 Methodology

### 3.1 Overview of the methodological approach

The proposed workforce planning framework consists of three main components: 1) input data/ feature engineering, 2) forecast engine, and 3) simulation component. Moreover, the forecasting horizon of the proposed framework is 6 years (2023 to 2028, 72 months ahead). In the first component, we focus on collecting data, selecting the top features from supply, demand, and other relevant variables.

This provides the inputs for the second component, where we develop forecasting algorithms and use the ensemble method to prepare the final forecast. This model offers *Integrated Care Board (ICB)* level projections with prediction intervals. In the final component, we test different scenarios to provide evidence for making policy-level decisions at the *National Health Service England (NHSE)* regional level. Figure 1 depicts an overview of the proposed workforce planning framework.

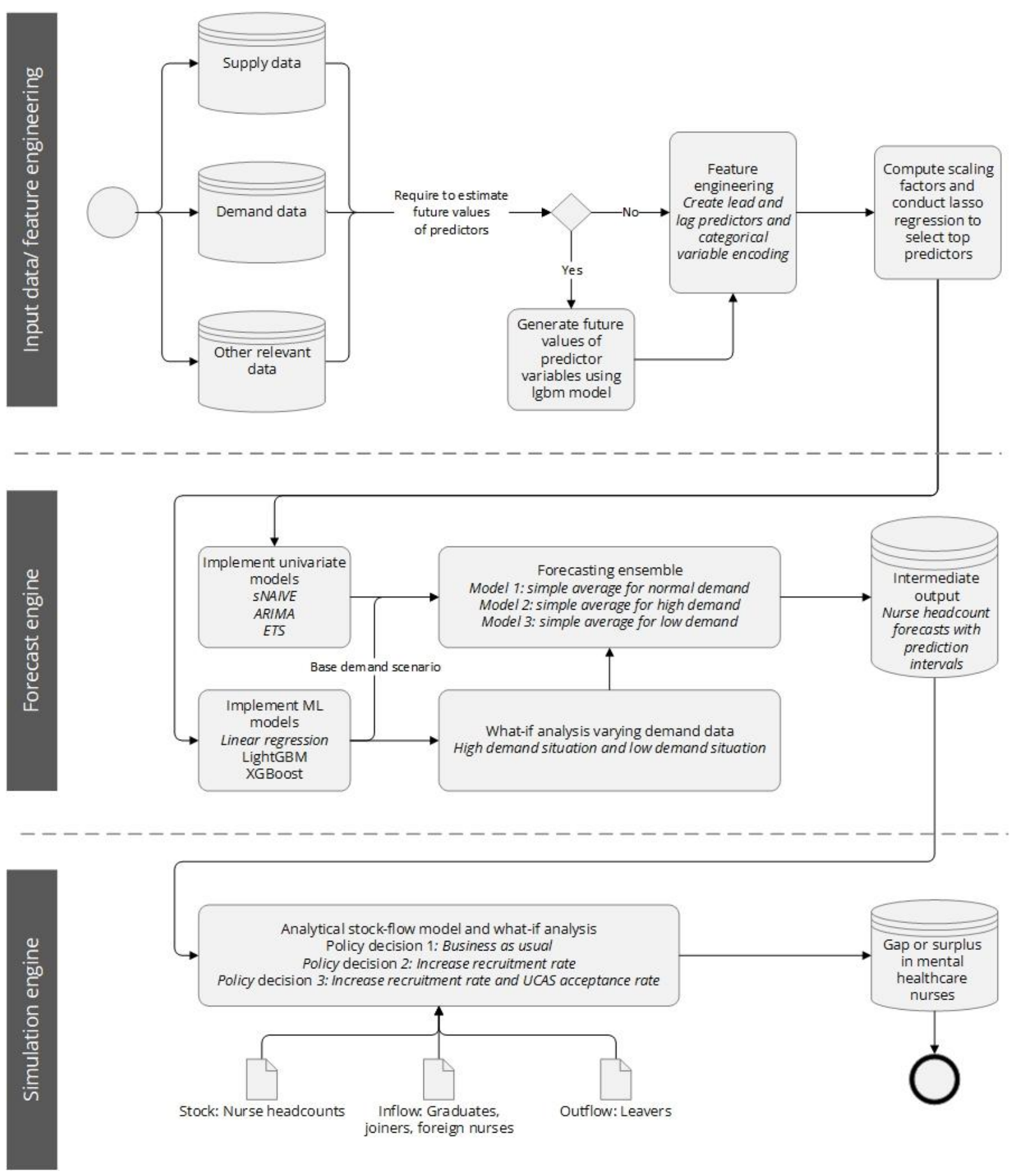

Figure 1: The proposed workforce planning framework consists of three components, namely 1) input data/ feature engineering, 2) forecast engine, and 3) simulation component.

### 3.2 Data sources and preprocessing

The datasets used and their corresponding sources are detailed in Table 2. To ensure uniformity in the time frame for our study, we opted for the period spanning June 2018 to December 2022. This decision was primarily driven by the fact that each source provided data covering varying time spans. Moreover, we focused on *Nurses & health visitors staff category* in *Mental Health and Learning Disability services* in our study.

Table 2: Data sources and descriptions.

| Source | Description | Period | Time granularity |
|---|---|---|---|
| NHS Workforce Statistics[1] | Headcounts of nurses and health visitors currently working | Jun 2018 – Mar 2023 | Monthly |
| | Leavers and joiner's headcount* | Jun 2018 – Mar 2023 | Monthly |
| NHS mental health dashboard[1] | Investments for mental healthcare | Jan 2017 – Aug 2023 | Quarterly |
| OCED health data[2] | Nursing graduates | 2010 - 2022 | Annually |
| | Foreign trained nursing count | 2011 - 2022 | Annually |
| Mental Health Services Monthly Statistics[1] | Number of people in contact with mental health services (MHS01) | Jan 2018 – Mar 2023 | Monthly |
| | Number of referrals (MHS32) | Jan 2018 – Mar 2023 | Monthly |
| | Number of people with an open hospital spell (MHS07) | Jan 2018 – Mar 2023 | Monthly |
| | Number of contacts with mental healthcare services (MHS29) | Jan 2018 – Mar 2023 | Monthly |
| NHS Sickness Absence Rates[1] | | Jan 2018 – Mar 2023 | Monthly |
| NHS Vacancy Statistics[1] | | Jan 2018 – Mar 2023 | Quarterly |
| NMC Registration data reports[3] | UCAS acceptance** for mental healthcare courses | Jan 2018 – Dec 2022 | Monthly |
| | Foreign nursing inflow into mental healthcare | Jan 2018 – Dec 2022 | Monthly |
| Population and migration[4] | Population projections for local authorities | 2010 - 2035 | Annually |

* Staff who leave a permanent or fixed-term contract are considered leavers, while staff who join a permanent or fixed-term contract are considered joiners

** UCAS acceptance refers to the number of successful admissions to higher education institutes

*Sources:* [1](NHS Digital, 2023), [2](OECD, 2023), [3](NMC, 2022), [4](ONS, 2023)

Our initial exploration of the datasets indicates the absence of missing values in each dataset. Regarding geographical levels, the dataset is composed of 7 NHSE regions, which are further subdivided into 45 ICB levels, resulting in a total of 52 timeseries. Subsequently, we merged all the datasets into a tidy dataset with 41 variables and 5,670 observations. This step was undertaken to streamline the modelling process in our study. Furthermore, we estimated monthly values using quarterly and annual data. We assumed that annual values are equally distributed among the months.

### 3.3 Data exploration

First, we explore the time series features of nurse headcounts. However, it is not visually feasible to plot all the individual series together to inspect the trend and seasonality simultaneously. Thus, we compute the time series features and present them in Figure 2, which illustrates that almost all the series exhibit low seasonality and strong trends.

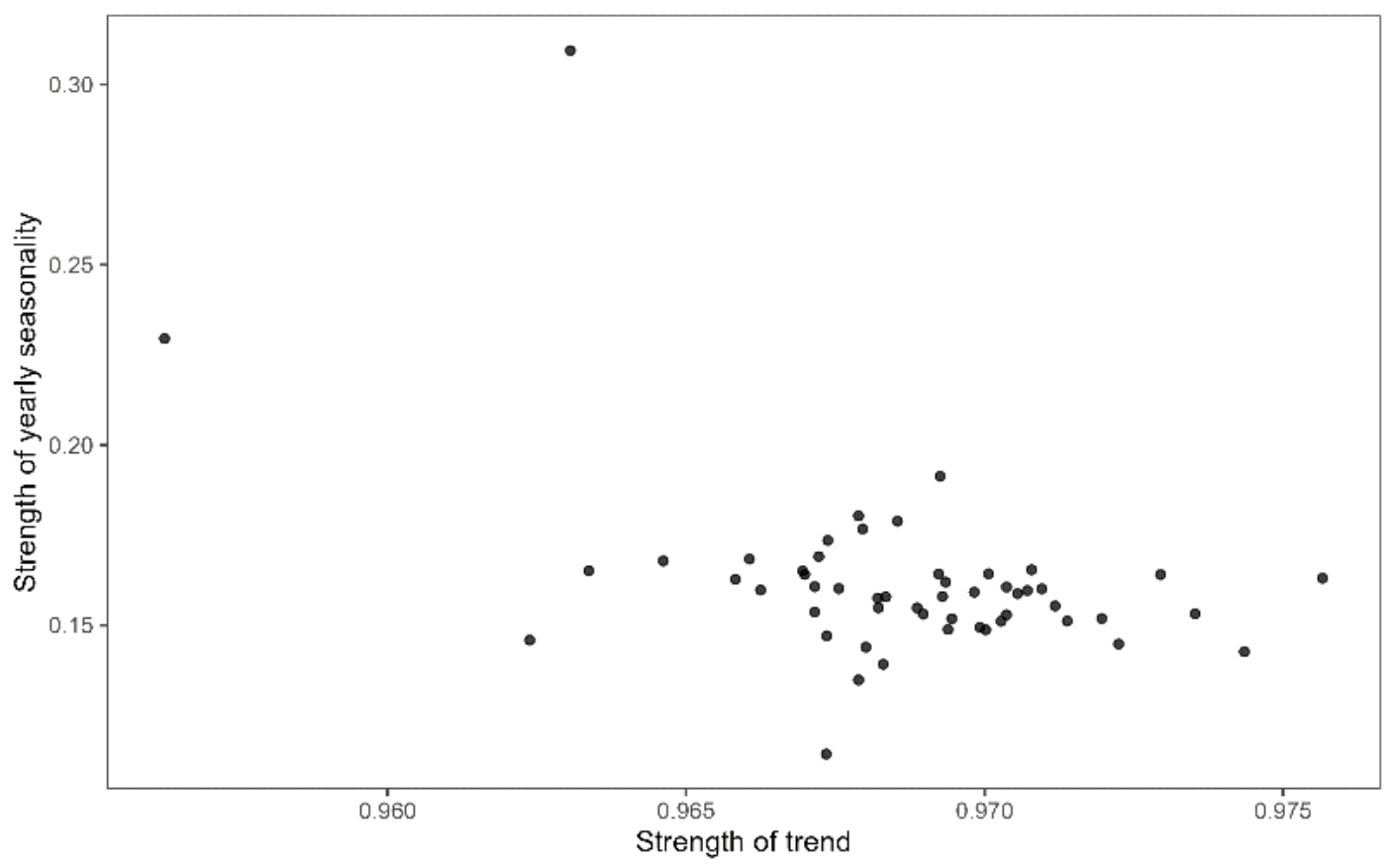

Figure 2: Timeseries features of nursing headcount.

Next, we examine the time series at various aggregation levels within the geographical hierarchy. Figure 3 reveals that although an overall growth of 4.74% (against June 2018) is observed at the higher level, indicating a general upward trend, differences emerge at lower levels. Notably, at the NHSE regional level, while most of the series exhibit a similar trend, Northwest shows a slightly decreasing trend. Moreover, at the ICB level, some ICBs experience an increasing trend, while others show a decreasing trend.

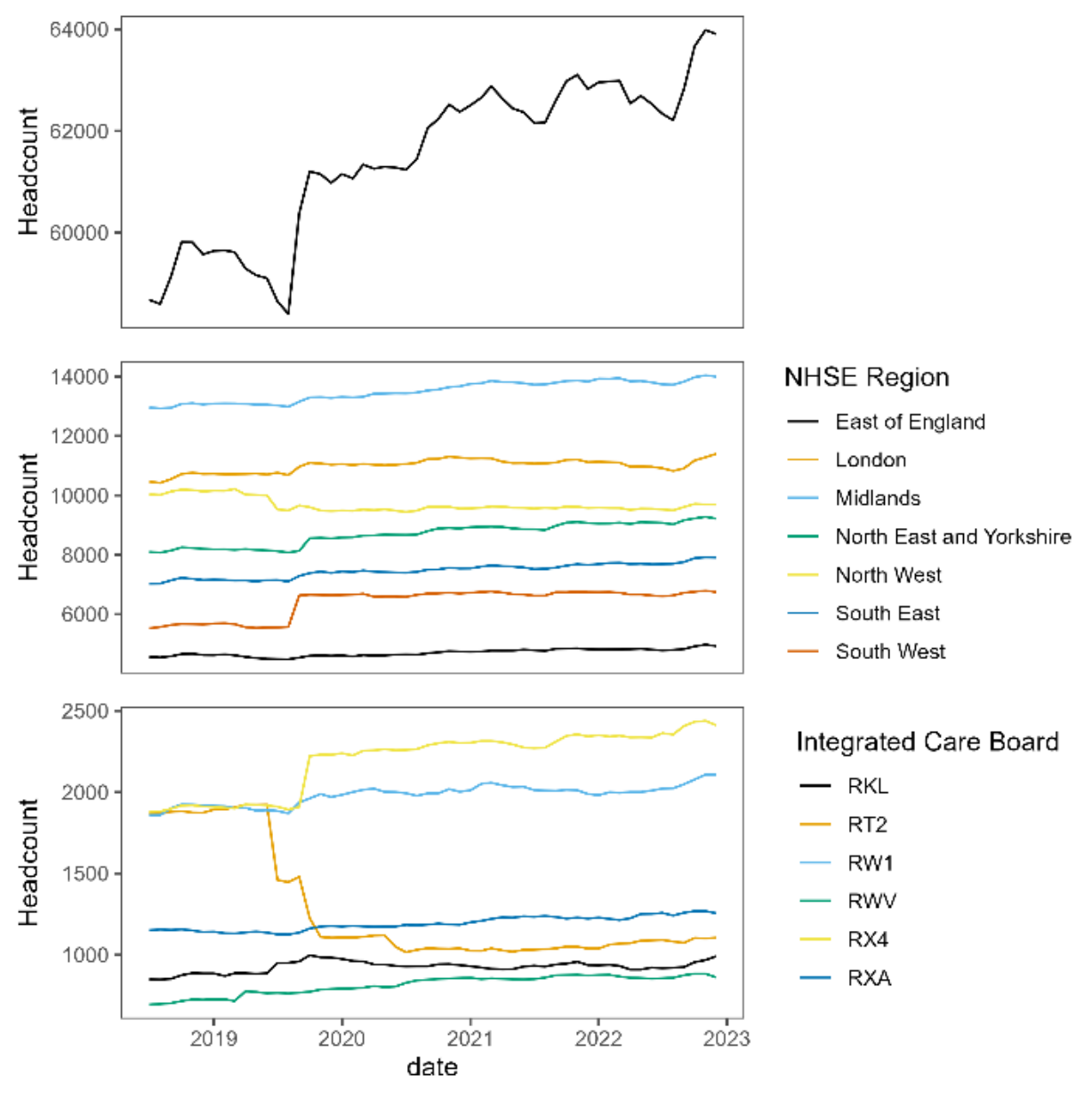

Figure 3: Timeseries of nursing headcounts in mental healthcare (June 2018 – December 2022).

*note: The three panels depict nursing headcounts from NHS England (first panel), organized by NHSE regions (second panel), and categorized by ICB levels (third panel). To ensure clarity and prevent overplotting, only 7 of the ICBs are shown.*

Particularly, certain series at the ICB level display high volatility at specific points in time, making the forecasting process challenging. To account for these diverse patterns in the time series at the ICB level, we consider several forecasting approaches in our study (see Section 3.5).

### 3.4 Input feature selection

We use a combination of causal and time series features. For causal variables, we consider supply, demand, and other relevant factors. Table 3 presents the comprehensive list of the driving factors that we incorporate into our study. To identify these factors, we consulted experts and referred to relevant literature (Anderson et al., 2021; Ono et al., 2013). Literature provides a general overview of driving factors applicable to workforce planning in various healthcare services. Additionally, for time series features, we utilize lag and lead predictors and date features for predictor variables, as outlined in (Kolassa et al., 2023). All the lag predictors we prepared are detailed in Table 3. To create lag features, we employed the *lag()* function in the "dplyr" package (Wickham et al., 2023).

Table 3: Target and predictor variables of our study

| Feature type | Description | Lag predictors |
|---|---|---|
| Target variable | Headcount of mental healthcare nurses | No |
| Demand factors | Number of people in contact with mental health services (MHS01) | Yes |
| | Number of referrals (MHS32) | Yes |
| | Number of people with an open hospital spell (MHS07) | Yes |
| | Number of contacts with mental healthcare services (MHS29) | Yes |
| | Population | No |
| Supply factors | Leavers headcount | Yes |
| | Joiners headcount | Yes |
| | Absence rate | Yes |
| | Number of vacancies | Yes |
| | UCAS acceptance rate | Yes |
| | International joiners in mental healthcare | Yes |
| | Nurse graduates (UK general) | Yes |
| | Foreign trained nurses (UK general) (SFTN) | Yes |
| | Foreign nurse inflow (UK general) (IFTN) | Yes |
| Other factors | Local spend on mental healthcare | No |
| | NHSE specialised commissioning spend on mental healthcare | No |
| | Total NHS spend on mental healthcare | No |
| | Clinical Commissioning Groups (CCG) spend on mental healthcare as a percent of base allocations | No |
| Static features | NHSE region | No |
| | ICB | No |
| Date features | Month | No |
| | Quarter | No |
| | Date | No |
| | Covid-19 indicator | No |

To select the most significant lag predictors, we explored the cross-correlation function (CCF) between the predictors and our target variable (i.e., the headcount of mental healthcare nurses). Subsequently, we employed lasso regression to identify the top predictors from the features pool. We used the lasso regression model as a global model, treating the data as cross-sectional. For the implementation of lasso regression in our study, we utilized the "sklearn" Python package (Pedregosa et al., 2011). Prior to fitting the lasso model, we used the *StandardScaler()* function to normalize the data. This step enabled us to automatically identify the most effective predictors for our modelling process.

We then generated future values for selected predictors, if estimation was necessary, using a simple LightGBM (LGBM) model (see Section 3.5 for more details). This model, in contrast to univariate models, can easily incorporate time series features like lags and rolling means into the forecasting process. Additionally, this model is efficient and straightforward to implement. We assume that historical trends and patterns will remain the same in the future when estimating these predictors. For population predictors, we used population and migration estimates provided by the Office for National Statistics (ONS, 2023).

We also created a COVID-19 indicator variable to capture the changes in trends caused by the pandemic. This variable is set to 1 from Q1 to Q2 2020 and 0 otherwise. The aim is to account for the shift in patterns that affected both supply (e.g., higher absence, increased leavers) and demand (e.g., more people seeking mental health support) during and after the pandemic. Including this variable helps the model better adjust for those changes when producing long-term forecasts.

### 3.5 Forecasting methods

As discussed in the previous section, we employ a range of forecasting toolbox from simple to complex in our study to account for the diverse patterns present in the nursing headcount. After generating these forecasts, we use an ensemble technique to produce the final forecast. For univariate methods, we utilize the sNAIVE, ARIMA, and Exponential Smoothing State Space (ETS) models. As machine learning models, we employ Linear Regression (LR), XGBoost (XGB), LGBM models. These models are efficient and simple to implement. In particular, the LGBM and XGB models efficiently handle large amounts of data (Januschowski et al., 2022). Importantly, all of these models have the capability to provide prediction intervals beyond just point forecasts. Therefore, we believe that incorporating these models into our framework enhances the practical implications in real-world scenarios.

***sNAIVE:*** We utilize this simple method as the initial model in our study. It involves generating forecasts using the last known observation from the previous period (Hyndman & Athanasopoulos, 2018).

***ARIMA:*** Widely applied in both industry and literature (Petropoulos & Svetunkov, 2020), these models utilize auto-correlation within time series data for forecasting. The ARIMA parameters (p, d, q) represent: p for auto-regressive (AR) component order, d for difference order, and q for moving average (MA) component order (Hyndman & Athanasopoulos, 2018).

In our study, we employed the *AutoARIMA()* function from the "StatsForecast" package in Python (Garza, Mergenthaler, et al., 2022). This function automatically fits multiple models by considering various p and q values and then selecting the best model using the lowest Akaike Information Criterion (AIC) (Hyndman and Athanasopoulos 2018).

***ETS:*** This model incorporates trends, seasonality, and error terms in time series data through approaches like additive, multiplicative, or mixed methods within the state space framework (Hyndman and Athanasopoulos 2018). These models can also automatically identify the best model using specified information criteria (Hyndman & Khandakar, 2008). To implement this model in our study, we utilized the *AutoETS()* function from the "StatsForecast" package in Python (Garza, Mergenthaler, et al., 2022).

***LR:*** Regression models establish linear relationships between the target variable and predictor variables. In our study, we used the *LinearRegression()* function from the "sklearn" package (Pedregosa et al., 2011), along with the *MLForecast()* function from the Python "MLForecast" package to implement the model (Garza, Canseco, et al., 2022).

***Gradient-Boosted Regression Trees:*** Over recent years, these methods have gained popularity in time series forecasting. Among them, LGBM and XGB implementations are prominent. These models have demonstrated significant performance, even in the recent M5 competition (Januschowski et al., 2022). Moreover, their increasing efficiency and simplicity make them well-suited for practical applications (Huber & Stuckenschmidt, 2020). To implement these models, we used *LGBMRegressor()* from the Python "LightGBM" package (Microsoft Corporation, 2022) and *XGBRegressor()* from the Python "xgboost" package (xgboost Developers, 2021), along with the *MLForecast()* function from the Python "MLForecast" package (Garza, Canseco, et al., 2022). Finally, we used grid search for the hyperparameter tuning for LGBM and XGB models in our study.

***Forecast ensemble:*** In time series forecasting, previous literature often reports that an ensemble of methods improves forecasting performance (Piccialli et al., 2021). The process of forecasting ensembling is also known as forecast combination (Godahewa et al., 2020). In ensembling, multiple forecasting models are usually trained independently and then aggregated together to produce the final forecasts (Timmermann, 2006). The aggregation method of these models is crucial for producing a better final forecast. The most common approaches for forecast combination are simple averaging and weighted averaging (Godahewa et al., 2020). In our study, we employed simple average approach to produce the final forecasts.

***What-if analysis:*** We conducted a what-if analysis to assess how variations in input values of demand-side factors impact the forecasting model. We examined three scenarios for the demand-side factors: business as usual, high demand, and low demand for mental healthcare services. This approach enables us to observe how the models respond to the introduction of uncertainty (Sutton et al., 2023), enhancing our capacity to adapt assumptions regarding demand-side factors and subsequently refine the forecasts (Marco et al., 2021).

### 3.6 Analytical stock-flow model

Simulation models have a substantial history of providing valuable support for strategic-level decision-making (Benahmed et al., 2023). Among these methods, analytical stock-flow models stand out as an efficient way to model the dynamics of the workforce system over time (Safarishahrbijari, 2018), and they are widely employed in healthcare workforce management (Ono et al., 2013). One of the advantages of analytical stock-flow models is that they are also capable of incorporating what-if analysis to test different policy decisions in the workforce system. Thus, this approach enables us to move beyond single-point forecasting or fully estimated probabilistic forecasting distributions by integrating forecasting with simulations. It empowers decision-makers to transform these forecasts, leading to a more profound understanding of the comprehensive healthcare workforce planning system (Anderson et al., 2021). Consequently, analytical stock-flow models have the capacity to bridge the gaps between supply and demand in healthcare workforce management. In our study, we used *AnyLogic software* to implement this model.

We employed the analytical stock-flow model on the supply side of the system to simulate the mental healthcare nurse supply. We defined the main inflows to the mental healthcare nurse stock are coming from mental healthcare nursing graduates, foreign nursing inflow, and other recruitments. Conversely, the primary outflow is the count of individuals leaving the profession. We used this model to estimate the supply of mental healthcare nurses and compared it against the output from the forecasting engine. We only utilized the base case demand estimates for mental healthcare nurses in the simulation model. Additionally, our consideration was limited to the NHS aggregate level and NHSE regional levels on an annual basis throughout the simulation process. The simulation runs from 2023 to 2028. Table 4 presents the initial values and assumptions of the stock-flow model.

Table 4: Initial values and assumptions of SD model.

| **Category** | **Variable** | **Initial value/ estimates** | **Assumption** |
|---|---|---|---|
| Inflow | Joiners headcount | Estimate using lgbm model for each region. | Historical trends and patterns remain the same. |
| | International joiners rate | 0.50% | NMC registry data and constant throughout the period. |
| | UCAS acceptance growth rate | East of England - 0.18<br>London - 0.03<br>Midlands - 0.27<br>North East and Yorkshire - 0.23<br>North West - 0.09<br>South East - 0.01<br>South West - 0.23 | Average annual growth rate calculated using NMC registry data. Constant throughout the period. |
| | UCAS acceptance initial value | East of England - 565<br>London - 990<br>Midlands - 1080<br>North East and Yorkshire - 905<br>North West - 970<br>South East - 505<br>South West - 325 | Number of successful admissions for the mental healthcare course in 2020 We assumed that the course period would be 3 years. |
| | Graduate joiners rate | 48% | NMC registry data. Constant throughout the period. |
| | Recruitment rate | 1 | This is set to 1 compared to the current recruitment rate. This will be an indication of how much to increase the current rate. |
| Stock | Headcount | East of England - 4850<br>London - 11062<br>Midlands - 13883<br>North East and Yorkshire - 9119 | NHS Digital. |

| Category | Variable | Initial value/ estimates | Assumption |
|---|---|---|---|
| | | North West - 9592 | |
| | | South East - 7758 | |
| | | South West - 6704 | |
| Outflow | Leavers headcount | Estimate using lgbm model for each region. | Historical trends and patterns remain the same. |
| Other | Demand forecasts | Output of the forecasting engine. | We only used the base case scenario. |

The stocks of mental healthcare nurses have initial values, which accumulate with total joiners and losses arising from leavers, as per the equations below:

$$Supply\ of\ nurses_{r,t} = INTEGRAL\left(total\ joiners_{r,t}\ -\ leavers_{r,t}\ , n_r\right) \quad (1)$$

where; $r$ is the NHSE region, $t$ is the time period and $n_r$ is the initial stock value of $r$ region.

$$Leavers_{r,t} = estimates\ from\ the\ lgbm\ model_{r,t} \quad (2)$$

The expected joiners come from mental healthcare nursing fresh graduates, international joiners, and other recruitments. Thus, we divided all joiners into these three categories, allowing us to control the variables. Moreover, we assume that fresh graduates join the mental healthcare service immediately after the completion of the program in the same NHSE region.

$$Growth_{r,t} = u_r * (1+v_r)^t * graduates\ joining\ rate \quad (3)$$

where;

$$Graduate\ joiners_{r,t} = u_{r,t-1} * (1+v_r) * graduates\ joining\ rate$$

where; $u_r$ is the initial number of successful admissions (UCAS acceptances) in region $r$ for year 0, $v_r$ is the annual growth rate of UCAS acceptance in $r$ region and $t$ is year.

$$All\ joiners_{r,t} =\ estimates\ from\ the\ lgbm\ model_{r,t} \quad (4)$$

$$International\ joiners_{r,t} = all\ joiners_{r,t} * international\ joiners\ rate \quad (5)$$

$$International\ joiners\ rate = 0.5 \quad (6)$$

$$Other\ recruitments_{r,t} = all\ joiners_{r,t} - graduate\ joiners_{r,t} - international\ joiners_{r,t} \quad (7)$$

$$\text{Total joiners}_{r,t} = \text{graduate joiners}_{r,t} + \text{international joiners}_{r,t} + \text{other recruitments}_{r,t} * \text{ recruitment rate} \quad (8)$$

Finally, we measure the supply and demand balance of mental healthcare nurses using the demand forecasts.

$$Demand\ for\ nurses_{r,t} = forecast\ engine\ output_{r,t} \quad (9)$$

$$Nurse\ supply\ \&\ demand\ balance_{r,t} = supply\ of\ nurses_{r,t}\ -\ demand\ for\ nurses_{r,t} \quad (10)$$

Figure 4 shows the stock-flow diagram for one region. The structure of the diagram is the same for the other regions. This measure was taken to ensure the legibility of the figure.

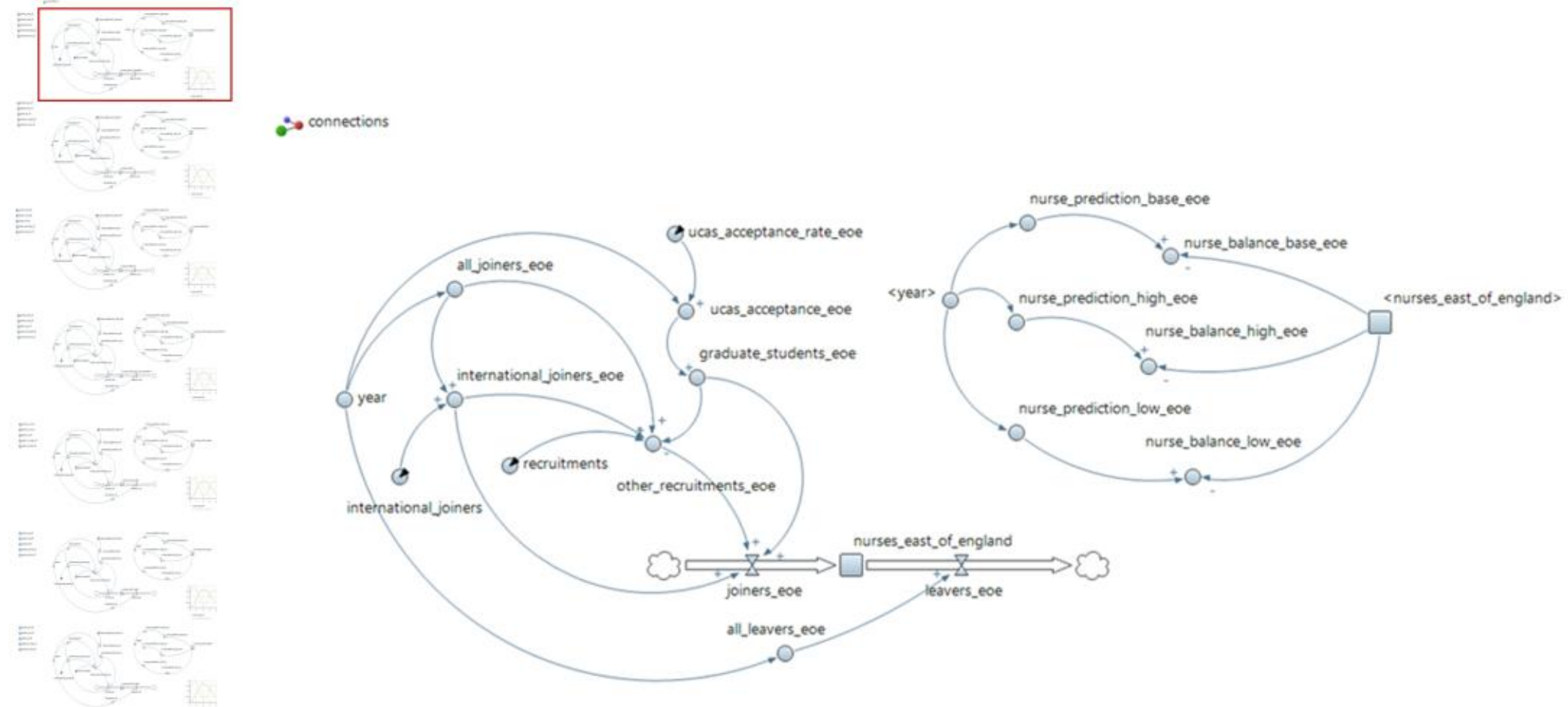

Figure 4: Proposed analytical stock-flow framework.

Next, we turn our attention to the what-if analysis of the proposed model. To conduct the policy analysis, we developed a systematic scenario analysis using a parameter grid. This grid varies four supply-side levers, UCAS acceptance growth rate, international joiner rate, recruitment multiplier, and graduate retention rate, across a plausible range of values informed by historical trends and expert discussions. Each combination of these values represents a unique scenario, enabling us to explore how these parameters interact and influence future nurse supply projections. Table 5 presents the structure of the parameter grid.

For each scenario in the grid, we apply the analytical stock-flow model across all NHSE regions over the forecast horizon (2023–2028). This setup allows us to assess the expected stock level in 2028 under each combination of assumptions.

Although this approach does not explicitly model external real-world drivers (e.g., changes in immigration policy or university funding), it offers a practical way to examine the implications of uncertainty in key controllable policy levers.

Table 5: Policy scenarios for the what-if analysis.

| Parameter | Description | Values tested |
|---|---|---|
| UCAS growth rate | Annual growth rate in UCAS acceptances | 0%, 10%, 20%, 30%, 40% |
| International joiner rate | % of international joiners from UCAS base | 0.1%, 0.3%, 0.5%, 0.7%, 1% |
| Recruitment multiplier | Relative increase in recruitment effort | 1.0, 1.1, 1.2, 1.3 |
| Graduate retention rate | % of graduates entering mental health roles | 40%, 45%, 50%, 55% |

## 3.7 Performance evaluation

In our study, we focus on long-term forecasting with a 6-year horizon (72 months). However, we only have 54 months of historical data (June 2018 to December 2022), which makes it difficult to use standard evaluation techniques like rolling-origin time series cross-validation. This is a common challenge in long-term forecasting, where the forecast horizon is longer than any realistic test window (Granger & Jeon, 2007). It is especially true in healthcare settings like mental health, where external factors, such as changes in behaviour or events like pandemics, can impact the system in ways that historical data cannot fully capture (Hovenga & Lowe, 2020). So, in our case, standard cross-validation methods, which are often used in short-term forecasting, are not suitable.

To deal with this, we used three strategies. First, we applied an ensemble forecasting method to reduce model and parameter uncertainty, which is widely supported in forecasting research (Godahewa et al., 2021). Second, we ran a set of what-if scenarios by changing the demand-side predictors to test how sensitive the forecasts are to changes in inputs. Third, and most importantly, we linked the forecasting model with a simulation model to carry out a policy analysis. This allowed us to see how forecast uncertainty might affect workforce decisions. We also worked closely with NHS experts, presenting our work to them, and incorporating their ongoing feedback throughout the model development process to ensure the methodological coherency. They helped us decide which data sources to use, what predictors to include, and how to interpret the outputs. While this approach does not replace a full accuracy evaluation, it gives a practical and transparent way to assess the usefulness of long-term forecasts when data is limited. Moreover, prior literature also adopts similar approaches to model evaluation in long-term forecasting settings (see, for instance, Hovenga and Lowe 2020, Ono et al. 2013, Willis et al. 2018).

# 4 Analysis and results

In this analysis, we explore four key areas. First, we examine the feature selection process and provide a summary of the key predictors influencing long-term nurse workforce needs in mental healthcare, based on the results of the CCF and lasso regression.

Second, we present the forecasted values for the future requirements of mental healthcare nurses. In this section, we concentrate on various time granularities such as months, quarters, and years, as well as hierarchical levels like NHS (total), NHSE regions, and ICBs. Additionally, we present the outcomes of the what-if analysis conducted on the forecasting framework. However, in this paper, we only report results at the NHS England aggregate level and the NHSE level.

Third, we delve into the results of the analytical stock-flow model. This model focuses on the supply side of mental healthcare nurses. Here, we compare the analytical stock-flow model's outcomes with the output of the forecasting framework to ascertain the balance between the supply and demand for mental healthcare nurses. Furthermore, we conduct a comparison of these results across different scenarios and evaluate the models.

## 4.1 Feature selection and preprocessing

To carry out feature selection, we first standardised all predictor variables and applied Lasso regression to the full training data, treated as a cross-sectional dataset. In our investigation, we set the parameter α to 0.1. The resulting $R^2$ value of the model was 62.77%. In addition to that we carried out parameter optimization to determine the optimal α value, and 0.1 yielded the most favourable $R^2$ value.

Consequently, guided by the outcomes of the lasso regression model, we selected the top 30 predictors for our study.

Figure 5 shows the top 30 predictors selected using Lasso regression. This step was used to identify a standardised set of predictors to be used across all forecasting models. Lasso was selected for this task because it provides a simple and consistent way to perform feature selection when dealing with a large number of correlated predictors. In particular, we treated all data as cross-sectional and standardised the features before applying Lasso. We do not interpret the coefficients from Lasso as feature importance scores. Instead, this step allowed us to reduce dimensionality and maintain a common feature set across both statistical and machine learning models.

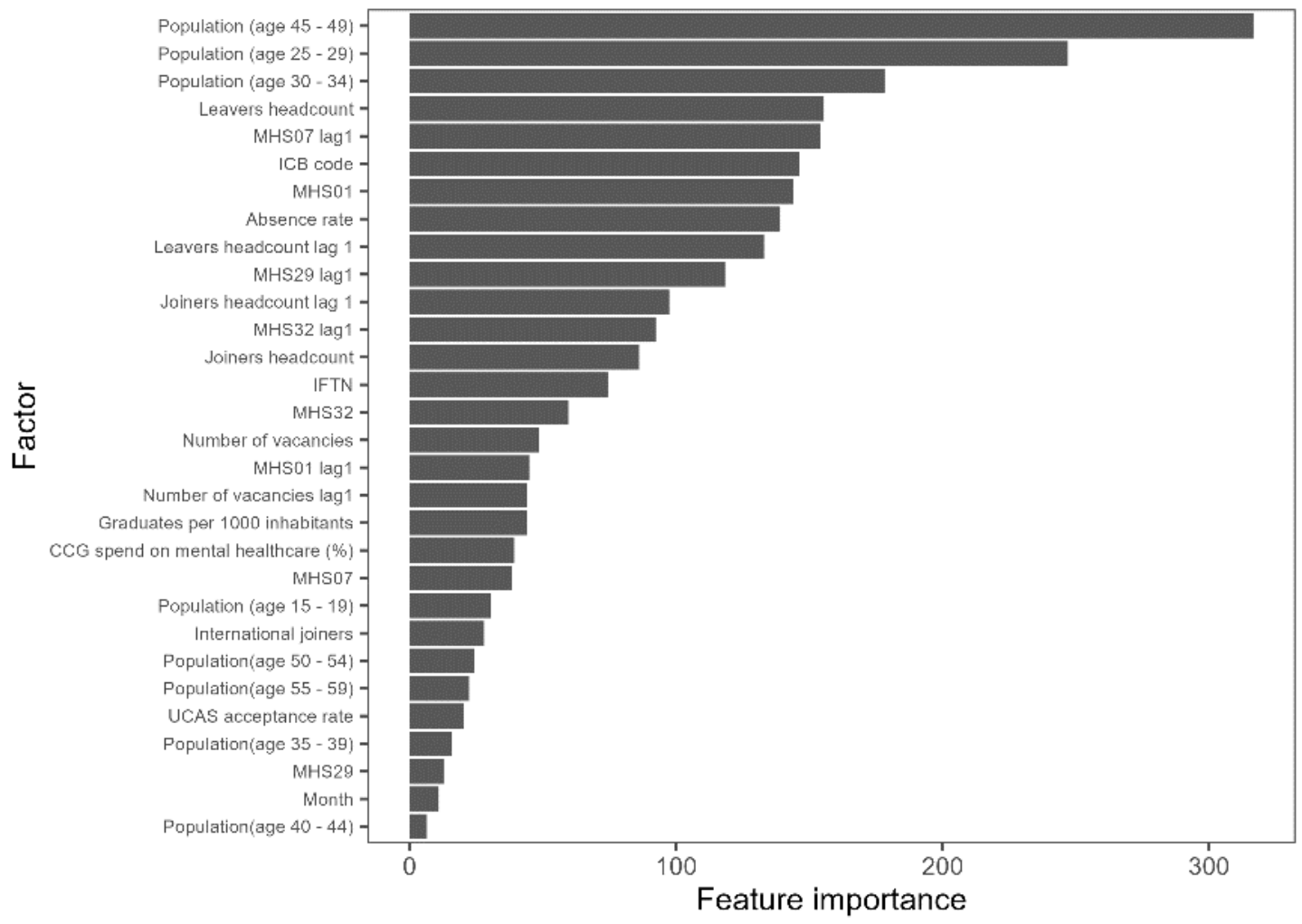

Figure 5: Feature importance of lasso regression.

Based on these findings, we finalized the subsequent predictors (as detailed in Table 6) for utilization within the forecasting engine.

Table 6: Finalised feature set for the model building.

| **Category** | **Feature type** | **Feature** |
|---|---|---|
| Supply based | Staff related | Leavers headcount and Leavers headcount lag 1 |
| | | Joiners headcount and Joiners headcount lag 1 |
| | | Absence rate |
| | | Number of vacancies Number of vacancies lag 1 |
| | | IFTN |
| | | International joiners |
| | Education and training | Graduates per 1000 inhabitants |
| | | UCAS acceptance rate |
| Demand based | Population demography | Population (age 15 – 19, age 25 – 29, age 30 – 34, age 35 – 39, age 40 – 44, age 45 – 49, age 50 – 54, age 55 - 59) |
| | Patient related | MHS01 and MHS01 lag 1 |
| | | MHS07 and MHS07 lag 1 |
| | | MHS29 and MHS29 lag 1 |
| | | MHS32 and MHS32 lag 1 |
| Other | Other relevant | CCG spend on mental healthcare (%) |
| | | ICB code |
| | Date features | Month<br>Covid-19 indicator |

## 4.2 Mental healthcare nurse demand forecasting results

### *4.2.1 Aggregate Demand for Mental Healthcare Nurses in NHS*

First, we examine the aggregate requirements for mental healthcare nursing. Figure 6 clearly illustrates an increasing trend in the demand for nurses throughout the forecasted period. On average, the demand for mental healthcare nurses is expected to increase by 3.52% in 2028 compared to 2022. The 95% prediction interval indicates the plausible range of future demand based on model and data uncertainty. Specifically, under the upper bound of this interval, demand may increase by up to 29.8%, whereas under the lower bound, it may decline by 22.8% compared to 2022. This interval reflects the uncertainty in future demand outcomes, not a confidence interval around the average forecast.

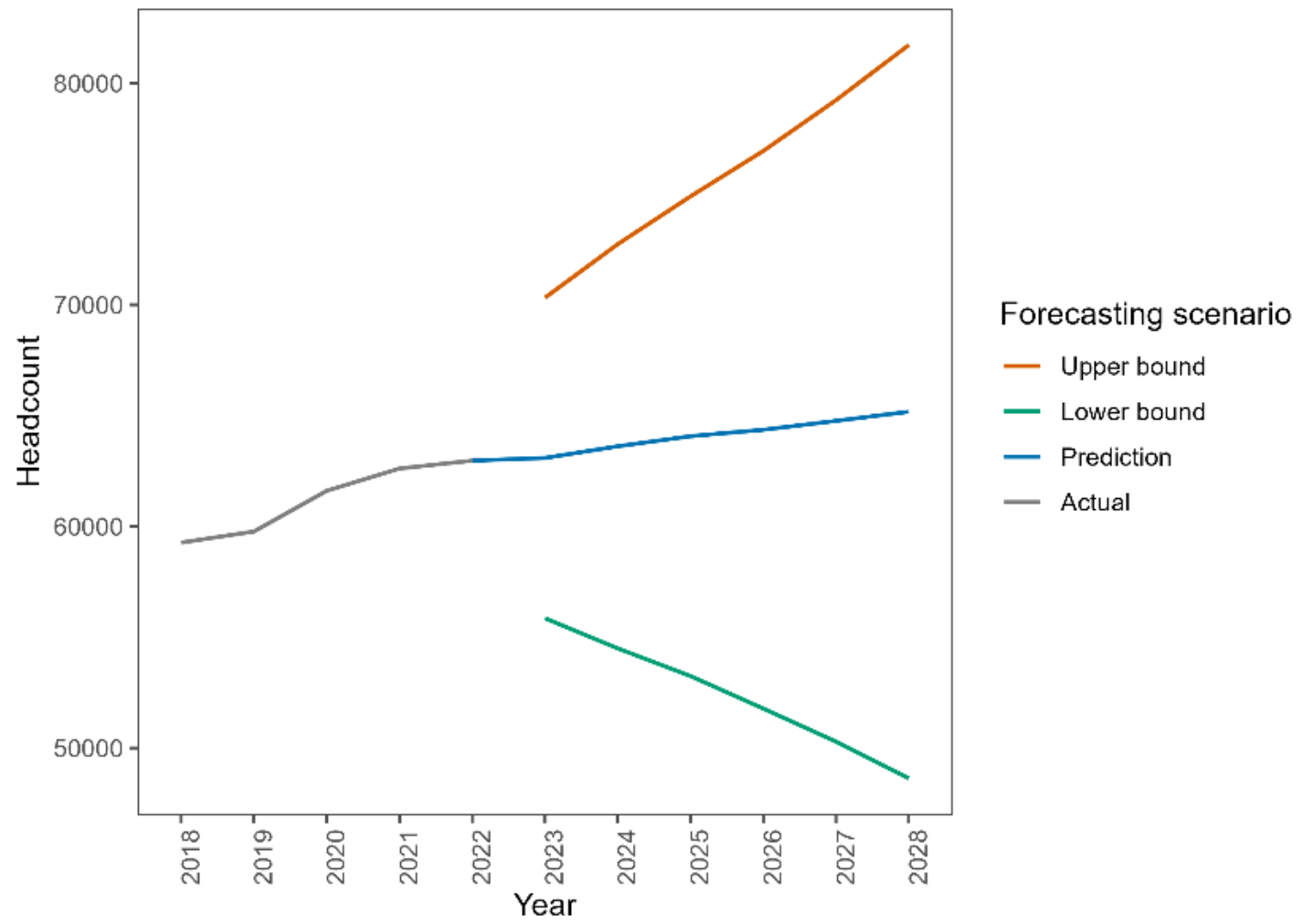

Figure 6: Annual predictions of demand for mental healthcare nurses at NHS level.

Table 7 and Figure 7 provide a clear depiction of the anticipated increase in demand for mental healthcare nurses across the majority of NHSE regions. Specifically, they reveal that London will experience the highest demand increase of 15% compared to 2022, while North East Yorkshire is projected to witness a decrease in demand by 1.51%. The Midlands and South East are projected to observe minor increases, with growth rates of 0.09% and 1%, respectively, indicating relatively stable demand growth. Moreover, South West, East of England, and South West regions are anticipated to have moderate increases in the demand for mental healthcare nurses. This highlights that while the overall demand for mental healthcare nurses grows at the aggregate level, regional-level requirements differ from one another.

Table 7: Growth rate of demand for mental healthcare nurses at NHSE regional level by 2028.

| Year | NHSE region | Prediction | Lower bound | Upper bound | Growth rate % (base) | Growth rate % (lower bound) | Growth rate % (upper bound) |
|---|---|---|---|---|---|---|---|
| 2028 | East of England | 5032 | 3916 | 6147 | 3.74 | -19.25 | 26.74 |
| 2028 | London | 12722 | 8433 | 17010 | 15.00 | -23.77 | 53.76 |
| 2028 | Midlands | 13895 | 12126 | 15663 | 0.09 | -12.65 | 12.83 |
| 2028 | North East and Yorkshire | 8981 | 7396 | 10567 | -1.51 | -18.90 | 15.88 |
| 2028 | North West | 9878 | 6031 | 13726 | 2.98 | -37.13 | 43.09 |
| 2028 | South East | 7835 | 6815 | 8856 | 1.00 | -12.16 | 14.15 |
| 2028 | South West | 6840 | 3923 | 9757 | 2.02 | -41.49 | 45.54 |

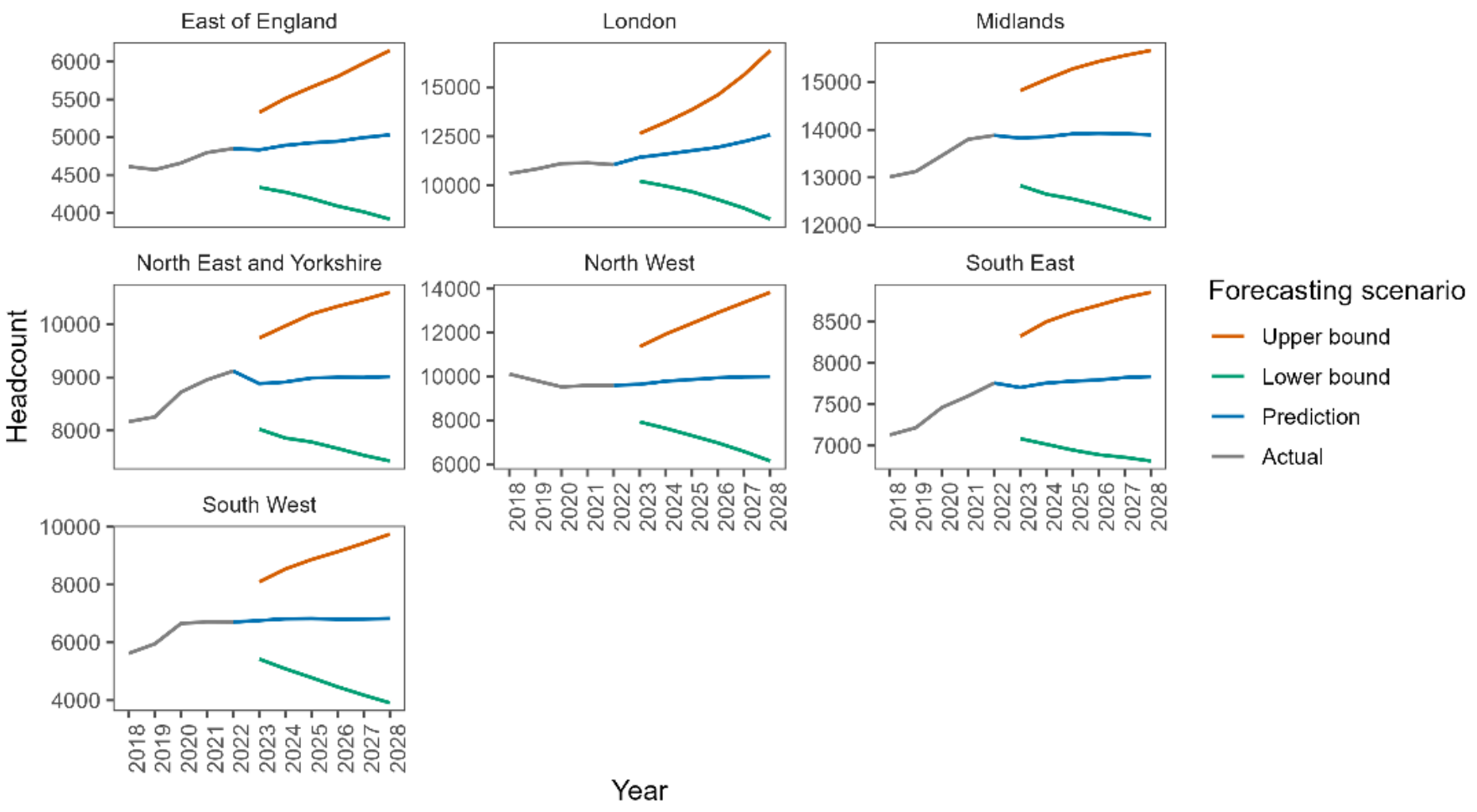

Figure 7: Annual predictions of demand for mental healthcare nurses at NHSE regional level.

.

### 4.3 What-if analysis of the forecasting engine

We evaluated the sensitivity of the forecasting engine by manipulating patient related predictors (specifically, MHS01, MHS07, MHS29, and MHS32). As detailed in Section 3.5, we formulated three scenarios for patient related estimations spanning from 2023 to 2028:

- **Base patient demand scenario:** Here, we forecast future values of patient-related predictors, assuming that historical trends and patterns will continue to hold true in the future.
- **High patient demand scenario:** In this scenario, we employed the upper bound estimates from the 95% prediction interval for patient related predictors.
- **Low patient demand scenario:** Conversely, for this scenario, we utilized the lower bound estimates from the 95% prediction interval for patient related predictors.

#### *4.3.1 What-if analysis on demand side at NHS aggregate level*

Figure 8 displays the prediction interval distributions for each scenario alongside the mean prediction. It demonstrates that the proposed forecasting model is sensitive to alterations in patient demand for mental healthcare and adjusts nurse demand accordingly for each scenario. Notably, the high demand scenario closely approximates the base scenario.

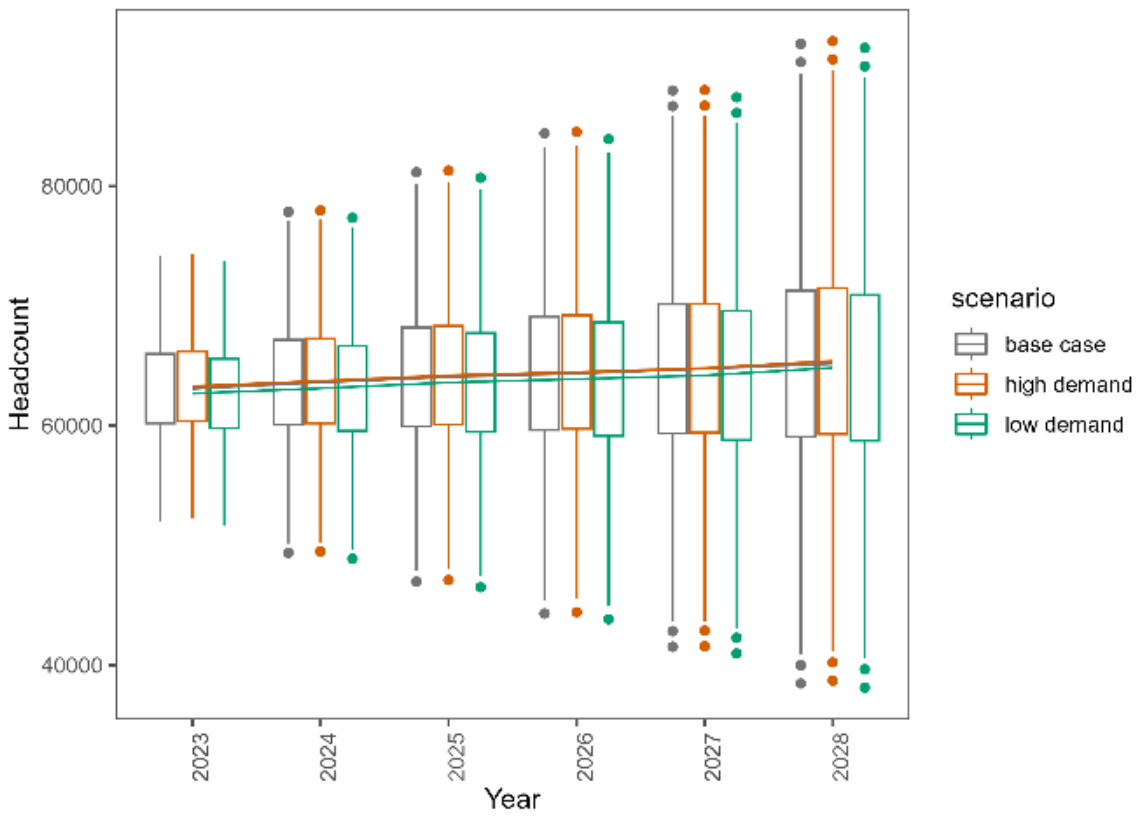

Figure 8: Annual predictions of demand distributions for each patient demand scenario at the NHS level.

This is further validated by Table 8. The table reveals an average 0.51% demand increase in the high patient demand scenario and an average 0.67% demand decrease in the low patient demand scenario, relative to the base scenario.

Table 8: Annual predictions of demand for each scenario at the NHS level.

| Year | Prediction (base case) | Prediction (high demand) | Prediction (low demand) |
|---|---|---|---|
| 2023 | 63057 | 63326 | 62688 |
| 2024 | 63530 | 63824 | 63114 |
| 2025 | 63983 | 64311 | 63587 |
| 2026 | 64280 | 64610 | 63840 |
| 2027 | 64604 | 64964 | 64197 |
| 2028 | 65183 | 65559 | 64683 |

### 4.3.2 *What-if analysis on demand side at NHSE regional level*

Figure 9 shows that the low patient demand scenario is always associated with generating less demand compared to the base demand for mental healthcare nurses across the regions. London stands out as experiencing the highest demand of 1.19% for mental healthcare nurses in the high patient demand scenario. However, the high patient demand scenario generates demand for mental healthcare nurses that are sometimes close to the base case scenario across the regions. Supporting this observation, Table 9 provides evidence. Specifically, North East & Yorkshire, North West, and South West exhibit relatively lower average demand for mental healthcare nurses in the high patient demand scenario, compared to the base scenario.

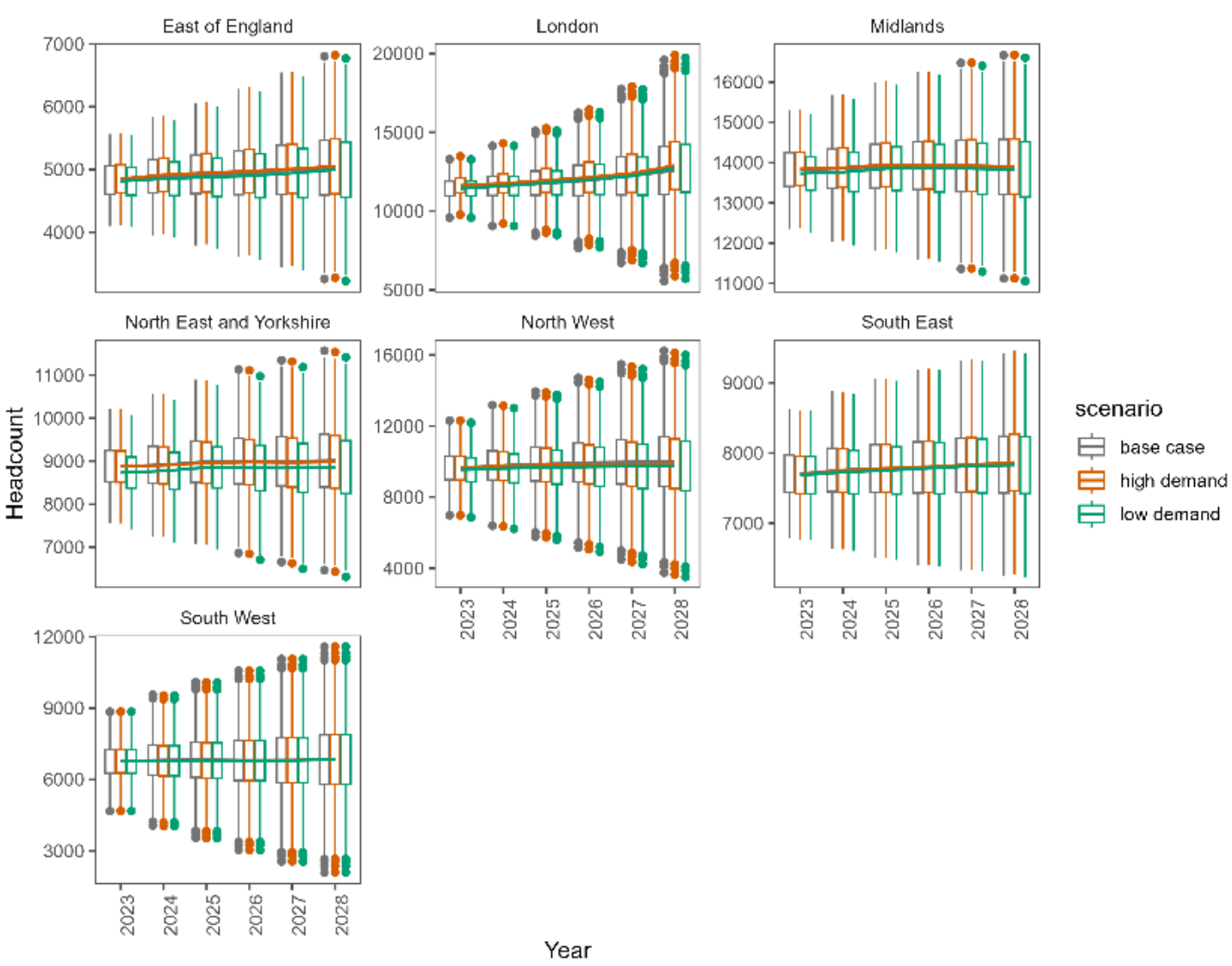

Figure 9: Annual predictions of demand distributions for each patient demand scenario at the NHSE regional level.

Table 9: Mean demand change for mental healthcare nurses for high and low patient demand scenarios compared to the base demand at the NHSE regional level.

| NHSE region | Prediction (base case) | Prediction (high demand) | Prediction (low demand) |
|---:|---:|---:|---:|
| East of England | 4925 | 4944 | 4890 |
| London | 11835 | 11977 | 11801 |
| Midlands | 13888 | 13902 | 13824 |
| North East and Yorkshire | 8969 | 8987 | 8860 |
| North West | 9765 | 9830 | 9659 |
| South East | 7776 | 7787 | 7764 |
| South West | 6787 | 6797 | 6784 |

## 4.4 What-if analysis on supply side using the analytical stock-flow model

### *4.4.1 Business as usual scenario*

First, we examine the aggregate level of mental healthcare nurse supply and demand. Figure 10 reveals that demand increases from 63,057 in 2023 to 65,183 in 2028, while supply falls from 63,184 to 59,914 over the same period. This indicates that the gap between supply and demand will continue to widen if the current recruitment approach remains unchanged. However, Figure 11 shows that the gap between supply and demand in different regions does not follow the same pattern as at the aggregate level. North East & Yorkshire and Midlands continue to maintain a surplus of mental healthcare nurses throughout the forecast period. Further, it shows that London and North West experience the highest increase in annual shortages by 2028, with deficits of 2,549 and 2,260 nurses respectively. Hence, it is evident that an examination of the NHSE region level is necessary to evaluate policy changes, as it offers a more holistic perspective.

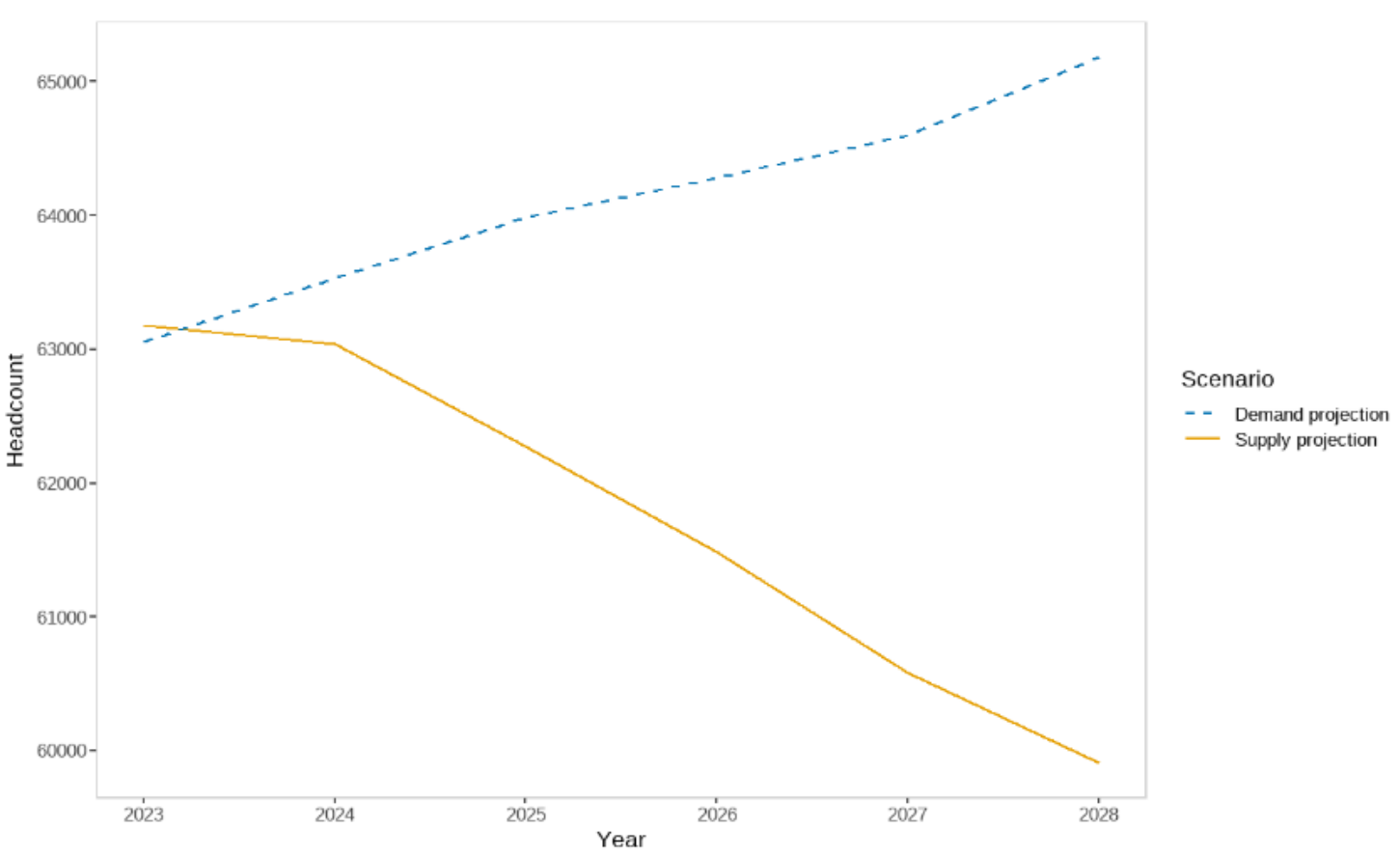

Figure 10: Supply projection and the demand projection for mental healthcare nurses at NHS level.

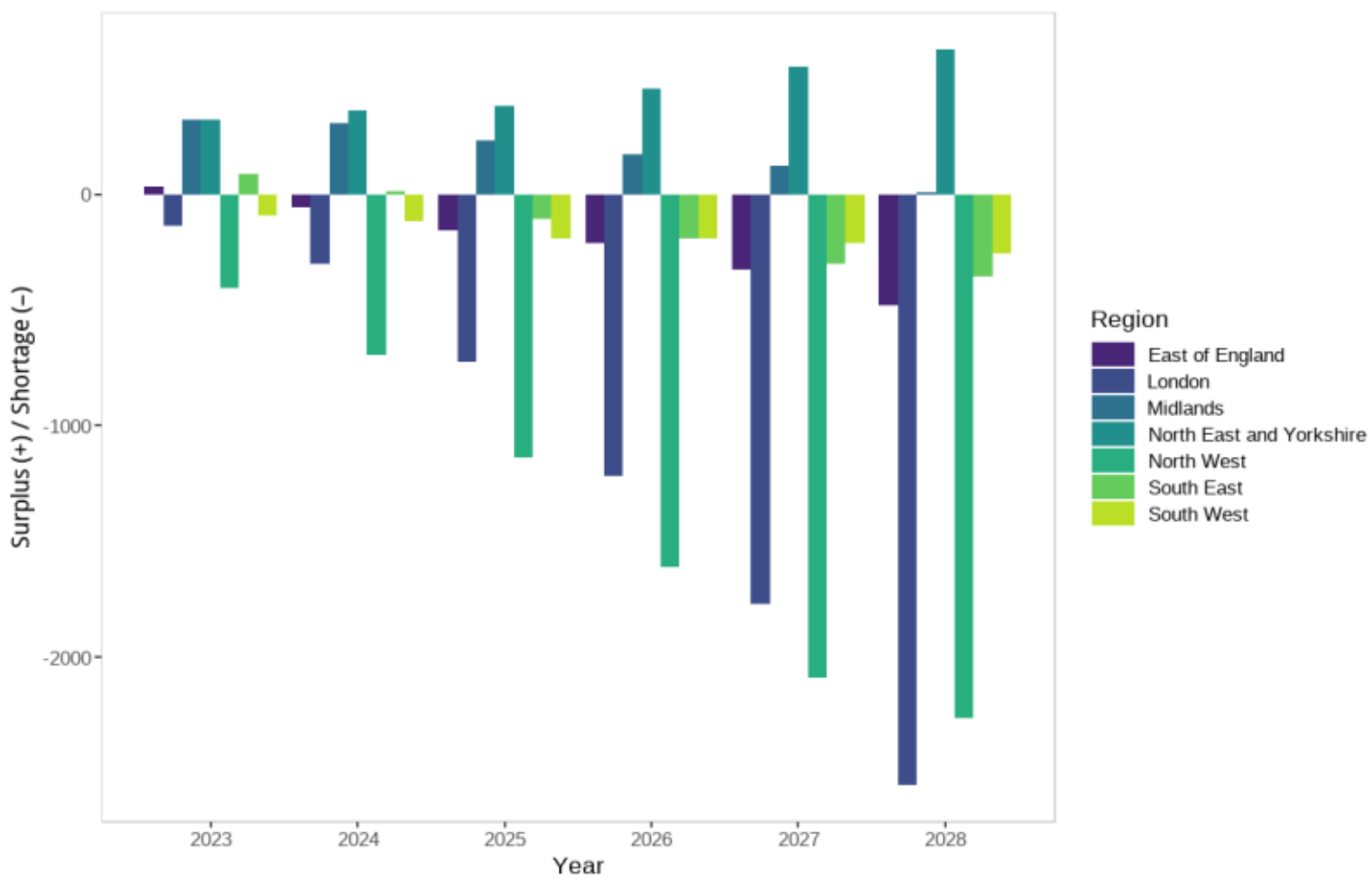

Figure 11: The gap between supply projections and the demand projections for mental healthcare nurses at NHSE regional level.

### *4.4.2 Policy Scenario Analysis*

We now examine the projected balance between mental healthcare nurse supply and demand under the full set of what-if scenarios generated using the parameter grid (see Section 3.6). Figure 12 presents the national-level fan chart, summarising the results from all scenario combinations. The chart shows the median projection (solid orange line), along with the interquartile (25%–75%) and 90% (5%–95%) ranges.

Under the majority of scenarios, the projected national supply exceeds demand by an increasing margin over time, indicating that the gap could be addressed through appropriate combinations of policy levers. However, the spread of the fan chart widens significantly over the forecast horizon, reflecting high uncertainty and model sensitivity to changes in UCAS growth, international recruitment, and graduate retention.

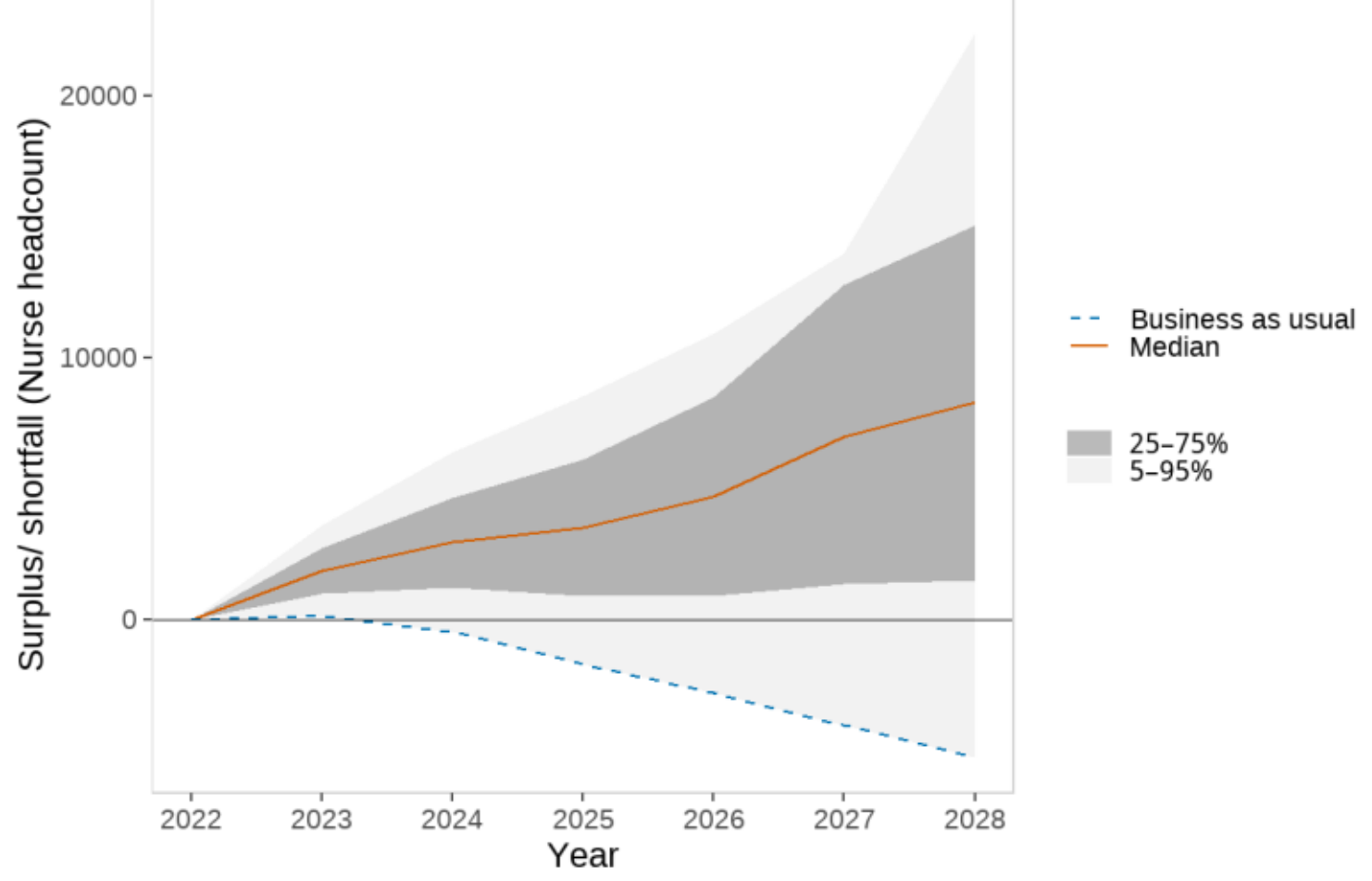

Figure 12: National-level forecast balance between mental healthcare nurse supply and demand across all what-if scenarios (2023–2028). The shaded areas represent the 25%–75% and 5%–95% ranges.

Figure 12 shows that the gap between supply and demand follows different patterns across NHSE regions. North East & Yorkshire and Midlands maintain a positive balance across most scenarios, including the business-as-usual case. This indicates that current workforce levels in these regions may remain sufficient if existing policies are sustained. However, regions such as London and North West demonstrate persistent shortages in nearly all scenarios. This suggests that these regions require stronger interventions to close the supply and demand gap. Hence, it is evident that a region-specific approach is necessary when designing workforce policies, as national-level policies may lead to over- or under-estimation of needs in specific regions.

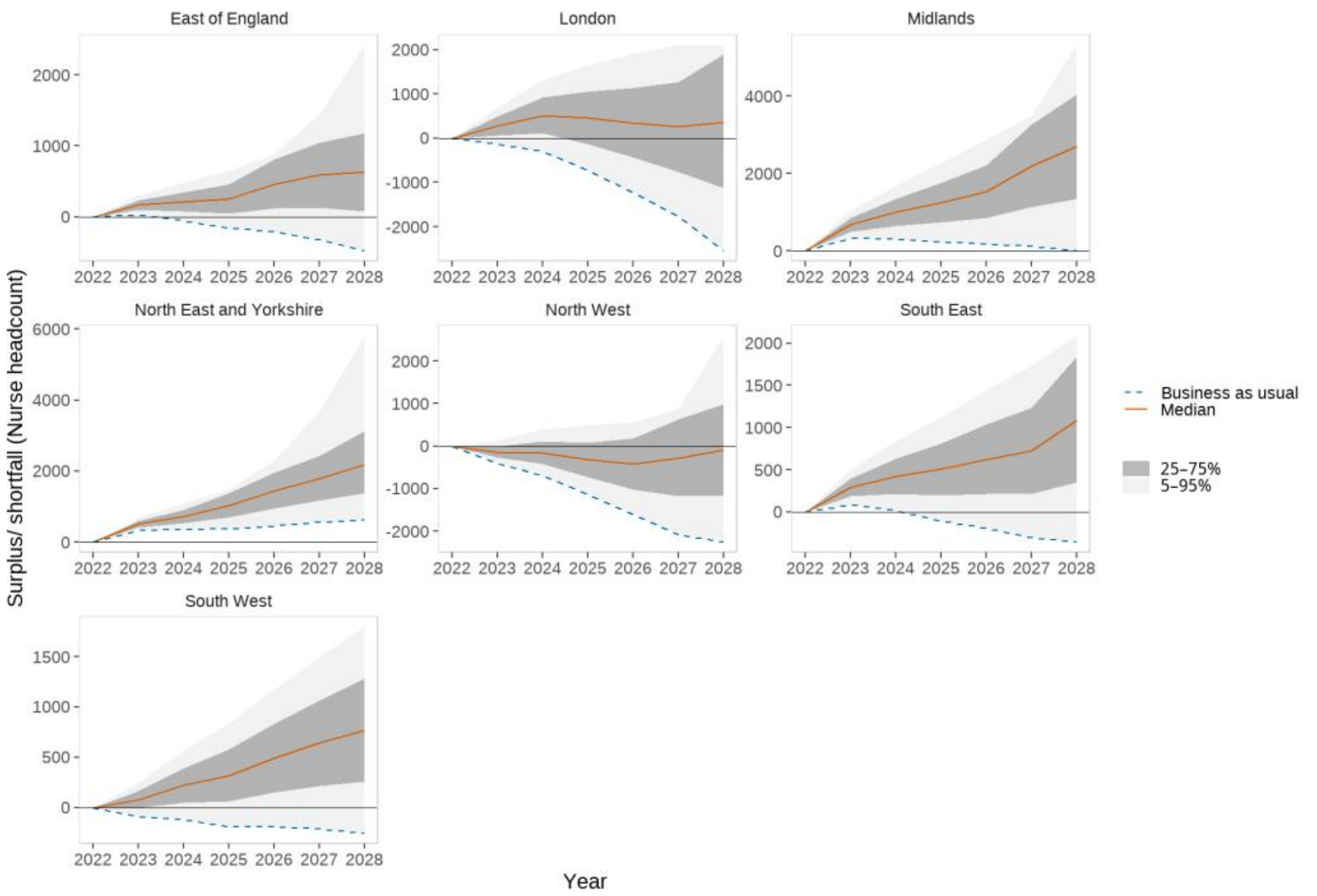

Figure 13: Regional-level forecast balance across all scenario combinations. Results show wide variation in projected surplus or shortage depending on policy mix and local context.

### *4.4.3 An illustrative example of alternative policy paths*

Next, we examine an example from the scenario analysis results for the London region. Table 10 presents three policy scenarios that all lead to an identical nurse balance of 2,090 by 2028. Despite achieving the same outcome, each scenario applies a different combination of policy levers.

For instance, Scenario A achieves the target by maintaining current UCAS acceptance levels and increasing graduate retention to 55%. In contrast, Scenario B applies a 20% increase in UCAS acceptances while keeping retention lower at 40%. Scenario C strikes a middle ground, with a 10% UCAS growth and moderate improvements in retention and international recruitment.

It shows that the same staffing outcome can be achieved through different routes, each with its own operational and political considerations. For example, increasing UCAS capacity requires long-term investment in nursing education, while improving retention may be more achievable in the short term through staff wellbeing initiatives or workplace reforms.

This example demonstrates that decision-makers can use the scenario grid not only to identify potential shortages but also to explore alternative and potentially more feasible policy pathways to close the gap. Further, it highlights the role of scenario analysis in identifying flexible and region-specific policy options. This is particularly important in practice, as policy feasibility, resource constraints, and implementation capacity vary across regions and over time.

Table 10: Illustrative scenarios for London region resulting in an identical 2028 nurse balance. While each scenario achieves the same outcome, the underlying policy levers differ, reflecting the flexibility of planning under uncertainty.

| Scenario | UCAS Growth | Int. Joiner Rate | Recruitment Multiplier | Graduate Retention | Nurse Balance (2028) |
|---|---|---|---|---|---|
| A | 0.0% | 0.005 | 1.3 | 55% | 543 |
| B | 20.0% | 0.003 | 1.3 | 40% | 543 |
| C | 10.0% | 0.007 | 1.3 | 45% | 543 |

# 5 Discussion

## 5.1 Findings

Long-term workforce planning in the healthcare sector is often compounded by various factors. Thus, preparing a reliable and robust workforce plan is always complex and challenging. However, policymakers always look for a reliable planning mechanism. Yet, many workforce planning methods in healthcare are unable to provide a holistic view as they often rely on either the supply or demand side of the process. Thus, there is a need for a model that captures all the related factors comprehensively and yet is simple enough to implement in the real world.

In this study, we proposed a hybrid predictive and prescriptive framework that brings together long-term forecasting with an analytical stock-flow model. By linking demand and supply projections, the model offers a structured way to explore how system behaviour might evolve under different future scenarios. Rather than offering fixed recommendations, our model is designed to facilitate scenario-based planning. It allows policymakers to test how changes in key inputs, such as recruitment, training, or service demand, can influence future workforce outcomes.

First, our findings show that several key driving factors influence the long-term demand for nurses in mental healthcare. These factors include both time series and exogenous factors. In terms of the exogenous factors, our study reveals that these include supply-based factors, demand-based factors, as well as other relevant factors. Concerning the supply-based factors, our study identifies the number of leavers and joiners, absence rates, the number of mental healthcare nurse graduates, foreign nurse inflow, and the number of vacancies as key drivers. On the other hand, in terms of demand factors, population demographics and the needs of different mental healthcare services have been identified. Additionally, factors such as investments in mental healthcare at regional and organizational levels have also been highlighted. These findings align with previous literature suggesting similar supply, demand, and other relevant factors to develop a robust workforce planning model (see, e.g., Anderson et al., 2021; Hovenga & Lowe, 2020; Ono et al., 2013; Willis et al., 2018).

Second, our findings show that incorporating exogenous features improves the forecasting process. For instance, during the what-if analysis process on the demand side predictors, when we alter the patient demand-related predictors, the forecasting engine responds to the changes and provides estimations accordingly. Thus, the addition of ML models improves the forecast engine's ability to detect these changes. This is also supported by previous literature; (Carter & Busby, 2023) explain that ML models have the capability to incorporate many features, often resulting in performance improvements when dealing with a large number of features. Moreover, this demonstrates how to incorporate uncertainty related to patient demand into the forecasting engine to capture the dynamics of the mental healthcare system. (Noorain et al., 2023) also explain that incorporating the uncertainty associated with patient-related factors into OR models is helpful for making informed management decisions in mental healthcare settings.

On the other hand, our study clearly demonstrates that each level of the organizational hierarchy shows different patterns of demand variations for mental healthcare nurses. This has also been highlighted by Bienemy (2015) and Ono et al. (2013), who stress the point that many projection models focus on the national level rather than considering geographical distribution. This may lead to inaccurate decisions. Our study provides evidence to support this. Noorain et al. (2023) also state that the demand for mental healthcare services varies across different locations and necessitates a model to aid decision-making across various planning levels. This is further supported by the results of the policy analysis. Our study's findings show that a "business as usual" approach is not sustainable for all regions. The results of the policy analysis clearly indicate that each region requires a tailored policy decision based on its specific characteristics. Therefore, our proposed model has the ability to capture demand variations at each level and provide reliable policy decisions using the hybrid predictive and prescriptive framework.

### 5.2 Managerial implications

A reliable workforce planning method is essential for decision-makers to ensure that people receive better and higher-quality mental healthcare services. In recent years, workforce planning methodologies have been evolving, especially with the emergence of simple and efficient quantitative methods like ML and simulation. However, decision-makers often continue to use fragmented planning approaches, which require substantial effort to convert outputs into actionable insights.

In this study, we propose a unified framework that brings together three key stages of the planning process, identifying key predictors, forecasting long-term workforce requirements, and evaluating policy options through scenario analysis. This integration supports planners to move beyond isolated projections and instead examine the combined impact of multiple demand and supply-side decisions in a structured and data-informed way.

The framework was specifically designed for practical use. It relies solely on publicly available NHS datasets and applies transparent, reproducible modelling techniques. Both the forecasting engine and simulation model are open and adaptable, allowing analysts to reflect local context and test region-specific assumptions. The model is not intended to provide prescriptive targets. Rather, it helps decision-makers explore different policy combinations and identify flexible strategies under uncertainty.

Initial feedback from the NHS Wales Modelling Collaborative and Health Education and Improvement Wales (HEIW) was positive. Both teams recognised the potential of the model to support long-term planning efforts and expressed interest in using the framework to inform strategic workforce discussions.

We emphasise that the outputs of the model are exploratory and scenario-based. For example, our analysis demonstrates that similar staffing outcomes can be achieved through different policy pathways, such as balancing higher UCAS intake with lower graduate retention, or vice versa. This reinforces a key point for decision-makers: there is no single solution to workforce shortages, and the ability to test alternative options is essential for developing resilient, evidence-based workforce strategies.

### 5.3 Limitations and Future Research Directions

While we have ensured the robustness of our study, a few limitations remain. Previous literature highlights the importance of more comprehensive and integrated models that focus on additional factors such as skill mix, working patterns, and demographic distribution of the workforce (Erhard et al., 2018; Kroezen et al., 2018). However, in our study, we did not incorporate skill mix and demographic distribution into the model due to limited data availability. For instance, we consider the total number of leavers from the mental healthcare workforce without distinguishing between categories such as retirement, career moves, or permanent transfers. This limits our ability to explicitly model the implications of an ageing workforce. A possible extension could involve configuring future scenarios in which the leaver trend is gradually increased to simulate higher exit rates due to retirement in specific regions. Exploring how to incorporate such demographic risk factors into the planning process would be an important avenue for future research.

Additionally, our model does not include factors such as productivity improvements resulting from operational and technological advancements. Safarishahrbijari (2018) explains that these factors can enhance the planning process and provide a new perspective on balancing the supply and demand of the healthcare workforce. Investigating how productivity improvement methods can address the supply and demand imbalance in healthcare workforce planning could be a valuable research topic. Moreover, wage improvements have been shown to positively impact nurse retention (Ono et al., 2013). However, due to limited data availability, we were unable to test such policy improvements in our model. Another way of incorporating factors that have limited data availability is to get expert support (De Bruecker et al., 2015). Incorporating such factors into the workforce planning model using expert input could be an area of exploration. Thus, studying how to integrate judgmental factors into quantitative-based healthcare workforce planning would be a worthwhile future endeavour.

In this study, we used Lasso regression for feature selection but did not explore model-specific interpretability techniques. Future work could investigate how individual features contribute to forecast accuracy using tools such as SHAP values or permutation feature importance. This would be especially useful for gaining further insight into the machine learning components of the ensemble.

Another limitation of this study is that we do not fully propagate the uncertainty from predictor forecasts into the final nurse headcount forecasts. While we use a what-if analysis to explore how changes in demand-side predictors affect forecast outcomes, our main prediction intervals are based on point estimates of the predictors. This means that the uncertainty bands shown in our results may underestimate the total forecast uncertainty. Incorporating a more formal uncertainty propagation step, for example, by sampling from the forecast distributions of the predictors and simulating multiple headcounts forecast paths, could be a valuable addition. Future studies could also explore probabilistic forecasting frameworks that account for parameter, model, and input uncertainty together.

Furthermore, our study only considers nurses and health visitors in the mental healthcare service at NHS England. Therefore, some findings, such as driving factors, may not be generalizable to other healthcare professions. Moreover, our model does not account for hierarchical reconciliation at local levels. We used a simple bottom-up approach to generate forecasts at different levels. However, our study reveals that each organizational level exhibits unique trends and patterns. Exploring hierarchical reconciliation of healthcare workforce forecasts by leveraging the hierarchical structure (e.g., geographical, or organizational levels) could be a promising avenue for future research.

Finally, our study relates to the assumption used to convert annual and quarterly data into monthly time series. For simplicity, we assumed that annual and quarterly values are evenly distributed across months. While this allowed us to create a consistent monthly dataset, it introduced a smoothing effect that may overlook meaningful intra-year variation. For example, nursing graduates tend to enter the workforce in specific months following graduation, rather than in a uniform monthly pattern. This assumption may limit the model's ability to capture seasonality or detect short-term shifts. Future work could explore more realistic disaggregation methods by using empirical evidence (e.g., graduation timelines or registry entry data) or drawing on expert knowledge to define more appropriate monthly distributions for such events.

### 5.4 Conclusion

The sustainability of a healthcare system relies on the reliability of its workforce. In light of the existing challenges in the mental healthcare workforce, it is crucial for healthcare organisations and policymakers to build a sustainable and sufficient supply of nurses to provide quality care. Therefore, we proposed the integration of a long-term forecasting model with a stock-flow model as a reliable tool for workforce planning in the mental healthcare nursing sector. The significance of taking regional levels into account when creating workforce estimates is emphasised by our research. Moreover, our analysis illustrates that maintaining the status quo in mental healthcare nurse workforce planning is not viable for all locations. Instead, alternative policy choices must be made, taking into account the unique features of each region. We believe that our suggested framework will provide decision-makers with evidence-based recommendations for developing mental healthcare nurse workforce policies that are well-suited to the mental healthcare requirements of the community.

**Data availability statement**

R and Python code to produce all the results in this paper is available at this [GitHub Repository]( https://github.com/chamara7h/longterm_workforce_planning).

**Disclosure statement**

The authors report there are no competing interests to declare.

**Acknowledgements**

We thank NHS Executive (NHS Wales) representatives Jennifer Morgan, Aura Frizzati, and Kate Burton for their enlightening conversations, which highlighted the crucial significance of building a model for mental healthcare workforce planning.